\documentclass[11pt,dvipsnames]{extarticle}

\usepackage{etoolbox}
\newcommand{\templatemode}{anonymous}   

\newif\ifanonymous   \anonymousfalse
\newif\iftsstyle     \tsstylefalse
\newif\ifpersonal    \personalfalse
\ifdefstring{\templatemode}{anonymous}{\anonymoustrue}{}
\ifdefstring{\templatemode}{tsstyle}{\tsstyletrue}{}
\ifdefstring{\templatemode}{personal}{\personaltrue}{}

\usepackage{amsfonts}
\usepackage{amsmath}
\usepackage{amssymb}
\usepackage{amscd}
\usepackage{amsthm}
\usepackage{mathrsfs}
\usepackage{graphicx}
\usepackage{wasysym}
\usepackage{xcolor}
\usepackage{tikz}
\usetikzlibrary{arrows.meta}
\usepackage{geometry}
\usepackage[12hr]{datetime}
\usepackage{textcomp}
\usepackage[utf8]{inputenc}
\usepackage[T1]{fontenc}
\usepackage[english]{babel}
\usepackage{appendix}
\usepackage{mathtools}
\usepackage{xfrac}
\usepackage{nicefrac}
\usepackage{url}
\usepackage{environ}
\usepackage[hang]{footmisc}
\usepackage[shortlabels]{enumitem}
\usepackage{multicol}
\usepackage{xpatch}

\usepackage{verbatim}
\usepackage{mdframed,lipsum}
\usepackage{fancyhdr}
\usepackage{rotating}
\usepackage{pdfpages}
\usepackage{setspace}
\usepackage{comment}
\usepackage{tabularx}

\iftsstyle
\usepackage{algorithm}
\usepackage[noend]{algpseudocode}
\else
\fi

\iftsstyle
\usepackage{authblk}
\fi

\iftsstyle
\usepackage{mathptmx}
\else\ifanonymous
\else
\usepackage[sc,osf]{mathpazo}
\usepackage[euler-digits,small]{eulervm}
\AtBeginDocument{\renewcommand{\hbar}{\hslash}}
\fi\fi

\iftsstyle
\usepackage[colorinlistoftodos]{todonotes}
\else

\fi

\iftsstyle
\PassOptionsToPackage{backref}{hyperref}
\fi
\usepackage{hyperref}
\usepackage[nameinlink, noabbrev, capitalize]{cleveref}
\crefname{appsec}{Appendix}{Appendices}

\iftsstyle
\hypersetup{colorlinks=true, linkcolor=blue, citecolor=violet}
\else
\hypersetup{colorlinks=true, urlcolor=NavyBlue, linkcolor=Mahogany,
	citecolor=ForestGreen, pdfborder={0 0 0}}
\fi

\iftsstyle
\else
\usepackage[
backend=biber,
style=alphabetic,
backref=true,
sorting=anyt,
maxalphanames=3,
minalphanames=2,
maxbibnames=99,
doi=false,
isbn=false,
url=false,
eprint=false
]{biblatex}

\ifanonymous

\else

\renewbibmacro{in:}{}
\DeclareFieldFormat[
article,inbook,incollection,inproceedings,
patent,thesis,unpublished
]{title}{#1}
\DefineBibliographyStrings{english}{
	backrefpage={},
	backrefpages={}
}

\DeclareFieldFormat{backrefparens}{\addperiod#1}
\xpatchbibmacro{pageref}{parens}{backrefparens}{}{}
\fi
\fi

\let\oldFootnote\footnote
\newcommand\nextToken\relax

\renewcommand\footnote[1]{%
	\oldFootnote{#1}\futurelet\nextToken\isFootnote}

\newcommand\isFootnote{%
	\ifx\footnote\nextToken\textsuperscript{,}\fi}

\newcommand{\dist}{\ensuremath{\operatorname{dist}}}

\newcommand{\tx}{\text}

\newcommand{\wgt}{\tx{\normalfont wt}}

\newcommand{\supp}{\tx{\normalfont supp}}

\newcommand{\sgn}{\tx{sgn}}

\newcommand{\wt}{\widetilde}

\newcommand{\poly}{\tx{\normalfont poly}}

\newcommand{\iref}[2]{(\hyperref[#2]{\ref*{#1}.\ref*{#2}})}

\newcommand{\diag}{\tx{\normalfont diag}}

\newcommand{\F}{\mathbb{F}}

\newcommand{\eps}{\varepsilon}

\newcommand\nodot[1]{}

\mathchardef\mhyphen="2D

\def\cA{\ensuremath{\mathcal{A}}}

\def\cC{\ensuremath{\mathcal{C}}}
\def\cD{\ensuremath{\mathcal{D}}}

\def\cS{\ensuremath{\mathcal{S}}}

\def\cV{\ensuremath{\mathcal{V}}}

\def\Z{\ensuremath{\mathbb{Z}}}

\newcommand\Fq{\F_q}

\def\dim{{\tx{\normalfont dim}}}

\usepackage{thmtools}
\usepackage{thm-restate}

\newtheorem{theorem}{Theorem}[section]

\newtheorem*{claim*}{Claim}

\newtheorem{proposition}[theorem]{Proposition}

\newtheorem{lemma}[theorem]{Lemma}
\newtheorem{corollary}[theorem]{Corollary}

\newtheoremstyle{TheoremNum}
{\topsep}{\topsep}              
{\itshape}                      
{}                              
{\bfseries}                     
{.}                             
{ }                             
{\thmname{#1}\thmnote{ \bfseries #3}}
\theoremstyle{TheoremNum}

\theoremstyle{definition}
\newtheorem{definition}[theorem]{Definition}

\iftsstyle
\numberwithin{equation}{section}
\fi

\makeatletter
\NewEnviron{restatethm}[2]{%
	\protected@write\@auxout{}{%
		\string\@restatetheorem{#1}{\detokenize\expandafter{\BODY}}%
	}%
	\begin{theorem}[#2]\label{#1}\BODY\end{theorem}%
}
\newcommand{\@restatetheorem}[2]{%
	\expandafter\gdef\csname restatethm@#1\endcsname{#2}%
}
\newcommand{\restatethmnow}[2]{%
	\begingroup
	\renewcommand{\thetheorem}{\ref{#1}}%
	\begin{theorem}[#2]\csname restatethm@#1\endcsname\end{theorem}%
	\endgroup
}
\makeatother

\makeatletter
\newcommand*{\transpose}{%
	{\mathpalette\@transpose{}}%
}
\newcommand*{\@transpose}[2]{%
	\raisebox{\depth}{$\m@th#1\intercal$}%
}
\makeatother

\title{}
\author{}
\date{}

\usepackage{booktabs,array}
\renewcommand{\wt}{\tx{\normalfont wt}}

\providecommand{\wR}{w_{\mathrm R}}
\providecommand{\dH}{d_{\mathrm H}}
\providecommand{\dR}{d_{\mathrm R}}
\providecommand{\rate}{\tx{\normalfont rate}}
\providecommand{\rank}{\tx{\normalfont rank}}

\providecommand{\YES}{\texttt{YES}}
\providecommand{\NO}{\texttt{NO}}
\providecommand{\Rat}{\operatorname{Rat}}

\providecommand{\muR}{\mu_{\tx{\normalfont R}}}

\providecommand{\Rep}{\operatorname{Rep}}
\providecommand{\notes}[1]{}

\title{On the NP-Hardness of Rank-Metric Minimum Distance and Syndrome Decoding, at Every Rate}
\author{Venkata Gandikota\thanks{Syracuse University}
\and Vatsal Jha\thanks{Purdue University}
\and S. Venkitesh\thanks{Ben-Gurion University of the Negev}}
\begin{document}

	\maketitle
	\begin{abstract}
		In this work, we investigate the hardness of computing minimum distance, as well as the hardness of syndrome decoding, for rank-metric codes.  It is known (Gaborit and Z\'emor, IEEE TIT 2016) that for \(\F_{q^m}\)-linear rank-metric codes of length \(N\), minimum distance when \(m=O(N)\) and syndrome decoding when \(m=O(N^2)\) are NP-hard under \emph{randomized} reductions.  We obtain a derandomization of these results, and a bit more.

		We show that for any fixed small \(\eps>0\), for \(\F_{q^m}\)-linear rank-metric codes of length \(N\) with \(m=O(N^{1+\eps})\), the problems of minimum distance, nearest codeword, and syndrome decoding are NP-hard under \emph{deterministic} reductions.  Further, we can prescribe the rate, that is, if we restrict to codes having any fixed rational rate \(R\in(0,1)\), then the NP-hardness for all three problems persists.

		Our main technical contribution is simple, and possibly of independent interest.  The reduction of Gaborit and Z\'emor (IEEE TIT 2016) relies on a distance-preserving embedding of a Hamming-metric code into a rank-metric code, obtained by multiplying its coordinates by a uniformly random vector.  We show that we can choose an explicit vector from a slightly larger extension field.
	\end{abstract}
	\clearpage

	\section{Introduction}
	\label{sec:intro}

	The theory of error-correcting codes is concerned with encoding messages in a structured way, so that they can be recovered even after being subject to noise. Two notions that are vital to such recovery are the \emph{redundancy} introduced by incorporating additional information derived from the message symbols into the codeword symbols, and the \emph{separation} between distinct codewords. The tradeoff between these two requirements is central to the study of error-correcting codes.

	\paragraph*{Codes and distance functions.}

	A code is a nonempty subset \(C\subseteq\Sigma^N\) over a finite alphabet \(\Sigma\), with a distance function on \(\Sigma^N\) representing the model of noise. One of the prototypical models uses the Hamming distance across coordinates. Even more structured is the setting of linear codes, where the alphabet is a finite field \(\F_q\). A linear code \(C\le\Fq^N\) is an \(\F_q\)-linear subspace, and its minimum distance \(\dH(C)\) is the least Hamming weight of a nonzero codeword. Computing \(\dH(C)\) is the \emph{minimum distance problem} (MDP). In the \emph{nearest codeword problem} (NCP), we instead seek a codeword closest to a given word. This is equivalent to \emph{syndrome decoding} (SD). These are basic computational problems in coding theory. Syndrome decoding was proved NP-complete in~\cite{BMT78}; the same was later proved for minimum distance in~\cite{Var97}. Subsequent work established NP-hardness of approximation within any constant factor under randomized~\cite{DMS03} and then deterministic reductions~\cite{CW12,AK14,BGR25}.

	Yet another important model is to expand codewords over an extension field to obtain a matrix over a base field, and replace the Hamming distance across coordinates by the rank distance.  The resulting codes are called \emph{rank-metric codes}, and these are the focus of our work. For \(x\in\F_{q^m}^N\), the rank weight is \(\wR(x)\coloneqq\dim_{\Fq}(\langle x_1,\dots,x_N\rangle_{\Fq})\), the rank of the \(m\times N\) matrix over \(\Fq\) obtained by coordinate expansion in a basis of \(\F_{q^m}\) over \(\Fq\). An \emph{\(\F_{q^m}\)-linear} rank-metric code is an \(\F_{q^m}\)-subspace \(C\le\F_{q^m}^N\). Introduced in~\cite{Del78,Gab85}, these are the rank-metric counterparts of Hamming linear codes. Gabidulin codes are the analogue of Reed--Solomon codes. They are used in network coding and form the basis of rank-based cryptography, whose security rests on rank syndrome decoding hardness.

	It must be noted that rank-metric codes are different from the related matrix codes. A \emph{matrix code} is an arbitrary \(\Fq\)-subspace of \(\Fq^{M\times N}\), for any number of rows \(M\). Placing a Hamming word on a matrix diagonal preserves its weight~\cite{Cou01}, so hardness in the Hamming metric implies hardness for matrix codes. The related \texttt{MinRank} problem is NP-complete~\cite{BFS99}. An \(\F_{q^m}\)-linear code must also be closed under multiplication by \(\F_{q^m}\), which the diagonal embedding need not ensure. The known hardness results for \(\F_{q^m}\)-linear rank-metric codes use the randomized reduction of~\cite{GZ16}: multiply the coordinates of a Hamming code \(C\) by \(\alpha\in(\F_{q^m}^*)^N\) and take the \(\F_{q^m}\)-linear span \(D_\alpha(C)_{\F_{q^m}}\). For uniformly random \(\alpha\) and \(m>2qN\), this code's rank distance equals \(\dH(C)\) with high probability~\cite[Thm.~8]{GZ16}. Their randomized reduction for rank syndrome decoding requires \(m>N^2\)~\cite[Thm.~10]{GZ16}. These reductions only exclude ZPP algorithms unless \(\tx{\normalfont NP}=\tx{\normalfont ZPP}\).
	We ask whether these reductions can be made deterministic.
	\begin{center}
		\emph{Is the minimum distance problem for \(\F_{q^m}\)-linear rank-metric codes NP-hard under deterministic reductions? Does this hold at every fixed rate?}
	\end{center}
	The first part was open even for exact minimum distance without parameter restrictions. We also prescribe the rate \(R=\dim_{\F_{q^m}}(C)/N\), which is fixed in applications such as rank-based cryptography. Our reductions produce hard instances at every rational rate \(R\in(0,1)\), while retaining positive relative distance and a threshold proportional to the length.

	\subsection{Our results}
	\label{subsec:results}

	\paragraph*{An explicit distance-preserving lift.} Our main technical result is a deterministic choice of the multiplier \(\alpha\) in the diagonal embedding of~\cite{GZ16}. It converts a Hamming-metric code into a rank-metric code with the same minimum distance. For \(C\le\Fq^N\) and \(\alpha\in(\F_{q^m}^*)^N\), define \(D_\alpha\coloneqq\diag(\alpha)\) and \(D_\alpha(C)\coloneqq\{\alpha\circ c:c\in C\}\), where \(\alpha\circ c\coloneqq(\alpha_1c_1,\ldots,\alpha_Nc_N)\). Write \(D_\alpha(C)_{\F_{q^m}}\coloneqq\langle D_\alpha(C)\rangle_{\F_{q^m}}\) for its \(\F_{q^m}\)-linear span. The reduction in~\cite{GZ16} chooses \(\alpha\) at random; we give an explicit choice.

	\begin{theorem}[Explicit diagonal lift]
		\label{thm:intro-lift}
		Let \(C\le\Fq^N\) be a nonzero linear code of dimension \(k\). Let \(\F_{q^m}\) be an extension field with \(m>k(N-k)\), let \(\vartheta\in\F_{q^m}\setminus\{0\}\) have degree \(m\) over \(\Fq\), and let \(\alpha=(1,\vartheta,\vartheta^2,\dots,\vartheta^{N-1})\). Then  \(\dR\bigl(D_\alpha(C)_{\F_{q^m}}\bigr)=\dH(C)\).
	\end{theorem}

	For every \(\alpha\in(\F_{q^m}^*)^N\), the inequality \(\dR(D_\alpha(C)_{\F_{q^m}})\le\dH(C)\) is immediate. Equality can fail for a bad choice of \(\alpha\). For instance, if \(\alpha\in(\Fq^*)^N\), the lifted code has rank distance~\(1\). Theorem~\ref{thm:intro-lift} gives an explicit choice of \(\alpha\) for which equality holds.

	\paragraph*{Deterministic hardness in the rank metric.} Theorem~\ref{thm:intro-lift} requires \(m\) quadratic in \(N\) when the rate is constant. Repeating every coordinate of the Hamming code before lifting keeps the dimension \(k\) fixed while multiplying the length. This lowers the required degree to \(m=O(N^{1+\eps})\). Applying the lift to any deterministic Hamming hardness result~\cite{AK14,BGR25} then gives the following.

	\begin{theorem}[Rank minimum distance]
		\label{thm:intro-rmdp}
		For every prime power \(q\), every \(\Gamma>1\) and every \(\eps>0\), it is NP-hard under deterministic polynomial-time reductions to distinguish \(\F_{q^m}\)-linear codes \(C\le\F_{q^m}^N\) with \(\dR(C)\le T\) from those with \(\dR(C)>\Gamma T\). This holds even when restricted to instances with \(m=O(N^{1+\eps})\), \(\dR(C)\ge\eta N\) and \(T\ge\theta N\), for constants \(\eta,\theta>0\) that depend only on \(q\), \(\Gamma\) and (if prescribed) the rate. Moreover, for every rational \(R\in(0,1)\), the code may be required to have rate exactly \(R\).
	\end{theorem}

	\begin{theorem}[Rank nearest codeword and syndrome decoding]
		\label{thm:intro-rncp}
		For every prime power \(q\), every rational \(R\in(0,1)\), every \(\Gamma>1\) and every \(\eps>0\), rank nearest codeword and rank syndrome decoding are NP-hard to approximate within gap \(\Gamma\) under deterministic reductions. This holds for instances of kernel rate exactly \(R\) with \(m=O(N^{1+\eps})\) and threshold \(T\ge\theta N\). The kernel code may also be required to have rank distance at least \(\eta N\), where \(\eta,\theta>0\) depend only on \(q,R,\Gamma\).
	\end{theorem}

	Theorem~\ref{thm:intro-rncp} replaces the randomized rank syndrome decoding reduction of~\cite{GZ16} by a deterministic one. For \(0<\eps<1\), its extension degree \(m=O(N^{1+\eps})\) also improves on their requirement \(m>N^2\). Their randomized minimum-distance reduction uses \(m=O(N)\).

	The formal problem definitions are in Section~\ref{sec:prelims}.

	\subsection{Technical Overview}
	\label{sec:intro:approach}

	Our reductions start from a known NP-hard problem in the Hamming metric and preserve the gap between \YES{} and \NO{} instances at each step. We describe the source problem, the four steps used to prove Theorems~\ref{thm:intro-lift}--\ref{thm:intro-rncp}, and how we combine them.

	\subsubsection*{The source problem}
	For the reductions below, we start from gap minimum distance in the Hamming metric. We state the source and derive its quantitative properties in Section~\ref{sec:body-source}.

	\begin{description}
		\item[(S) Hamming source.] For every \(\Gamma>1\) there is a deterministic polynomial-time reduction from 3SAT to pairs \((C,T)\), where \(C\le\Fq^n\) is a nonzero linear code and \(T\) is an integer, such that
		\[
		\YES\implies\dH(C)\le T,\quad\NO\implies\dH(C)>\Gamma T.
		\]
		Every output satisfies \(\dH(C)\ge\eta_0n\) and \(T\ge\theta_0n\), where \(\eta_0,\theta_0>0\) depend only on \(q,\Gamma\).
	\end{description}

	The hardness of the source problem (S) follows from the results of~\cite{AK14,BGR25}.

	For the decoding problems, we also need a low-weight witness whose value at a specified coordinate is \(1\). The reduction of~\cite{BGR25} provides this additional property.

	\begin{description}
		\item[(S\('\)) Distinguished-coordinate source.] For every \(\Gamma>1\) there is a deterministic polynomial-time reduction from 3SAT to triples \((V,j,T)\) with \(V\le\Fq^n\), integer \(T\), and \(j\in[n]\), such that
		\[
		\YES\implies\exists v\in V:\ v_j=1,\ \|v\|_0\le T,\quad\NO\implies\forall v\in V\setminus\{0\}:\ \|v\|_0>\Gamma T.
		\]
		Every output satisfies \(T\ge\theta_0n\) and \(\|v\|_0\ge\eta_0n\) for all \(v\in V\setminus\{0\}\), where \(\eta_0,\theta_0>0\) depend only on \(q,\Gamma\). The code \(V\) may be zero in the NO case.
	\end{description}

	\subsubsection*{Four steps}

	The lift (P1) preserves minimum distance, as stated in Theorem~\ref{thm:intro-lift}. Coordinate repetition (P2) lowers the required extension degree, and rate calibration (P3) fixes the rate exactly. Finally, (P4) converts a distinguished-coordinate source to a decoding instance. We show that combining these steps preserves the gap, positive relative distance, and a threshold proportional to the length.

	\paragraph*{(P1) The explicit diagonal lift (Hamming to rank distance)}
	For a nonzero code \(C\le\Fq^N\) of dimension \(k\), use \(\alpha=(1,\vartheta,\dots,\vartheta^{N-1})\), where \(\vartheta\ne0\) has degree \(m>k(N-k)\). Theorem~\ref{thm:intro-lift} gives \(\dR(D_\alpha(C)_{\F_{q^m}})=\dH(C)\), preserving thresholds and gaps. The proof has three steps.
	\begin{itemize}
		\item \emph{Step 1.} By Lemma~\ref{lem:rat}, the required lower bound is equivalent to the following condition for every \(U\le\Fq^N\) with \(r=\dim_{\Fq}(U)<\dH(C)\):
		\[
		D_\alpha(C)_{\F_{q^m}}\cap U_{\F_{q^m}}=\{0\}\iff\rank_{\F_{q^m}}[D_\alpha G\mid B]=k+r.
		\]
		Here \(G\in\Fq^{N\times k}\) and \(B\in\Fq^{N\times r}\) have columns forming bases of \(C\) and \(U\) (Lemma~\ref{lem:intersection}). A random choice handles each \(U\) separately, followed by a union bound~\cite{GZ16}; our choice handles all \(U\) at once.
		\item \emph{Step 2.} For \(D(t)\coloneqq\diag(1,t,\ldots,t^{N-1})\), a top-degree argument gives \(D(t)C_{\Fq(t)}\cap U_{\Fq(t)}=\{0\},\quad\rank_{\Fq(t)}[D(t)G\mid B]=k+r.\) A nonzero intersection would give \(0\ne c\in C\) with \(\dH(C)\le\|c\|_0\le r<\dH(C)\), a contradiction. Thus \(P_R(t)\coloneqq\det([D(t)G\mid B]_R)\ne0\) for some \(R\subseteq[N]\) with \(|R|=k+r\) (Lemma~\ref{lem:symbolic}).

		\item \emph{Step 3.} Write \(P_R(t)=t^aQ_R(t)\), where \(0\ne Q_R(t)\in\Fq[t]\) and \(\deg_t(Q_R(t))\le k(N-k)<m\). Thus \(Q_R(\vartheta)\ne0\), and \(P_R(\vartheta)\ne0\) since \(\vartheta\ne0\) (Lemma~\ref{lem:spread}).
	\end{itemize}

	The embedding \(C\mapsto D_\alpha(C)_{\F_{q^m}}\) is that of~\cite{GZ16}. (P1) provides an explicit multiplier to derandomize their construction. The strategy of replacing a random point by an indeterminate, proving nonvanishing over \(\Fq(t)\), and substituting an element of large degree is the usual way to derandomize a Schwartz--Zippel argument. The complete proof of Theorem~\ref{thm:intro-lift} is presented in Section~\ref{sec:lift}.

	\paragraph*{(P2) Coordinate repetition (lowering the extension degree)}
	By Lemma~\ref{lem:elementary}, repeating each coordinate \(L\) times gives \((n,k,d,T)\mapsto(nL,k,Ld,LT)\), preserving the gap and relative distance. For an integer \(B\ge1\), set \(L=n^{B-1}\) and \(N=n^B\). Then \(k\le n\) gives \(k(N-k)\le nN=N^{1+1/B}\).

	Coordinate repetition preserves the gap and appears, for example, in reductions between metrics~\cite{BRWW22}. Here, it increases the length while preserving the dimension. Since the degree bound in (P1) depends on both, repetition lowers the exponent in that bound.

	\paragraph*{(P3) Rate calibration (fixing the rate exactly)}
	For a rank-metric code \(C\le \F_{q^m}^n\), write \(R=a/b\) in lowest terms. Choose an explicit Gabidulin code \(D\) with \(\dR(D)\ge n\) and rate above \(R\), long enough that \(z\coloneqq b\cdot\dim(C\oplus D)-a\cdot\operatorname{length}(C\oplus D)\ge0\). Then
	\[
	\rate\bigl((C\oplus D)^{\oplus a}\oplus0^z\bigr)=\frac{a\cdot\dim(C\oplus D)}{a\cdot\operatorname{length}(C\oplus D)+z}=\frac ab.
	\]
	For nonzero \(C\), its minimum distance is \(\min\{\dR(C),\dR(D)\}=\dR(C)\), since \(\dR(C)\le n\). The construction requires \(m\ge\kappa_R n\) for a constant \(\kappa_R\) depending only on \(R\). We include this linear lower bound when choosing the field degree for the lift.

	For a decoding instance with target \(u\), use the same constructed code and target \(((u,0),0,\ldots,0)\). Only the first copy of \(C\) is translated. Projection onto the original coordinates and extension by zeros show that the minimum weight in the resulting affine space is unchanged. The distance bound for the auxiliary code also preserves a positive relative distance for the decoding code. Section~\ref{body:calibration-section} proves these assertions, including the case \(C=0\).

	\paragraph*{(P4) The affine slice (from minimum distance to decoding)}
	Given a source \((V,j,T)\) as in (S\('\)), first test \(A\coloneqq\{v\in V:v_j=1\}\) for emptiness. If \(A=\emptyset\), output the fixed NO instance constructed in Section~\ref{body:hardness-section}. Otherwise choose \(a\in A\), and put \(V_0\coloneqq\{v\in V:v_j=0\}\). Then \(A=a+V_0\) and \(\dist(-a,V_0)=\min_{v\in A}\|v\|_0\). In the \YES{} case, \(A\) contains a word of weight at most \(T\). In the \NO{} case, every word of \(A\) is a nonzero element of \(V\), so it has weight greater than \(\Gamma T\). Thus \((V_0,-a,T)\) has the same gap, and Lemma~\ref{lem:ncp-sd} gives syndrome decoding. The same argument works in the rank metric after (P1).

	Passing between minimum distance and nearest codeword by fixing one coordinate is the coding analogue of the reduction from SVP to CVP in~\cite{GMSS99}. This reduction replaces a basis vector \(b_i\) by \(2b_i\) and uses \(b_i\) as the target, so a closest vector corresponds to a short vector with an odd \(i\)-th coefficient. The embedding in~\cite{Kan87} goes the other way. For codes, the distinguished-coordinate version that yields a single Karp reduction is given in~\cite{BGR25}.

	For minimum distance, we apply coordinate repetition to the ordinary Hamming source, then the lift, and finally rate calibration. For decoding, we repeat and lift the distinguished-coordinate source, fix the corresponding coordinate to its required nonzero value, and then calibrate the linear code of differences. Section~\ref{body:hardness-section} verifies both reductions and their parameter bounds.

	Figure~\ref{fig:dependencies} records these dependencies and the order of the constructions.

	\begin{figure}[t]
		\centering
		\begin{tikzpicture}[
			>=Stealth, x=1cm, y=1cm,
			box/.style={draw, rounded corners, align=center, font=\small,
				inner sep=5pt, text width=5.3cm, minimum height=1.1cm},
			src/.style={box, fill=black!6},
			mid/.style={box, dashed},
			lab/.style={font=\small, align=center, fill=white, inner sep=3pt},
			every edge/.style={draw, ->}
			]
			\node[src] (S) at (0,0)
			{Ordinary Hamming source (S)\\Lemma~\ref{body:source}};
			\node[src] (SP) at (6.8,0)
			{Distinguished-coordinate source (S\('\))\\Lemma~\ref{body:source}};

			\node[mid] (LC) at (0,-2.3)
			{Lifted code \(\cC\coloneqq D_\alpha(C^{(L)})_{\F_{q^m}}\)\\
				\(\dR(\cC)=L\cdot\dH(C)\)};
			\node[mid] (LV) at (6.8,-2.3)
			{Lifted code \(\cV\coloneqq D_\alpha(V^{(L)})_{\F_{q^m}}\)\\
				\(\dR(\cV)=L\cdot\dH(V)\)};

			\node[mid] (A) at (6.8,-4.6)
			{Affine slice\\
				\(\cA\coloneqq\{x\in\cV:x_{j_\star}=\alpha_{j_\star}\}\)};

			\node[box] (RMD) at (0,-6.9)
			{Rank MDP at rate \(R\)\\Theorem~\ref{thm:intro-rmdp}};
			\node[box] (RNC) at (6.8,-6.9)
			{Rank NCP/SD at kernel rate \(R\)\\Theorem~\ref{thm:intro-rncp}};

			\path (S) edge node[lab] {(P2), then (P1)} (LC)
			(SP) edge node[lab] {(P2), then (P1)} (LV)
			(LC) edge node[lab] {(P3): homogeneous calibration\\
				Lemma~\ref{body:calibration}} (RMD)
			(LV) edge node[lab] {(P4)} (A)
			(A) edge node[lab] {(P3): affine calibration} (RNC);
		\end{tikzpicture}
		\caption{Dependencies of the rank-metric hardness results. Shaded boxes are the ordinary and distinguished-coordinate sources derived in Section~\ref{sec:body-source}; dashed boxes are intermediate instances. (P1) is the explicit diagonal lift (Theorem~\ref{thm:intro-lift}); (P2) is repetition before lifting (Lemma~\ref{body:repetition}); (P3) is rank calibration using a Gabidulin auxiliary code (Lemma~\ref{body:calibration}); and (P4) is the affine slice. Corollary~\ref{body:exact-lift} combines the three steps on the left. The right branch shows the nonempty-slice case: we lift the source code before fixing the coordinate and calibrate the direction code of \(\cA\). An empty source slice gives the fixed calibrated NO instance of Section~\ref{body:hardness-section}. For repeated length \(N_1\), we choose \(m\ge\kappa_RN_1\) to support calibration. For each rational \(R\in(0,1)\) and fixed \(\eps>0\), the final code or decoding kernel has rate exactly \(R\), distance at least \(\eta N\), and threshold \(T\ge\theta N\), with \(m=O_{R,\eps}(N^{1+\eps})\) for final length \(N\). Here \(\eta,\theta>0\) depend only on \(q,R,\Gamma\). Lemma~\ref{lem:ncp-sd} gives the syndrome form.}
		\label{fig:dependencies}
	\end{figure}

	\subsection{Related work}
	\label{sec:intro:related}

	\paragraph*{Concurrent work.}  The current manuscript is under submission.  We recently noticed a concurrent work~\cite{gaborit2026maximumlikelihood} that achieves deterministic reduction via a different idea, and in fact, for the stronger parameter setting \(m=O(N)\).  While we focused on derandomizing the specific reduction in~\cite{GZ16}, the concurrent work~\cite{gaborit2026maximumlikelihood} sets up a completely different deterministic reduction, also of independent interest.

	\paragraph*{Hamming metric.} Syndrome decoding was shown to be NP-complete in~\cite{BMT78}, and minimum distance in~\cite{Var97}, using a Reed--Solomon code concatenated with a Hadamard code. NCP was proved NP-hard to approximate within any constant factor via PCPs in~\cite{ABSS97}. The analogous result for MDP was proved under randomized reductions, using locally dense codes, in~\cite{DMS03}, which also introduced the relatively near codeword problem. The locally dense code construction was derandomized in~\cite{CW12}. A simple deterministic reduction through tensor codes was given in~\cite{AK14}; its hard instances have rate and relative distance bounded below by constants. The binary case was simplified in~\cite{Mic14}. PCP-free deterministic APX-hardness of NCP and MDP was established in~\cite{BGR25}. This argument is incorporated in~\cite{BGLR25}, which also gives PCP-free randomized hardness for sparsest vector problems. We develop our Hamming source from~\cite{BGR25} in Section~\ref{sec:body-source}; either~\cite{AK14} or~\cite{BGR25} suffices for the minimum-distance hardness.


	\paragraph*{Rank and matrix-rank metrics.} Rank-metric codes were introduced in~\cite{Del78}, and independently in~\cite{Gab85,Rot91}. \texttt{MinRank} for general matrix spaces is NP-complete~\cite{BFS99}, and the diagonal embedding transfers Hamming hardness to matrix codes~\cite{Cou01}. For \(\F_{q^m}\)-linear codes, the only previous hardness result is the randomized reduction of~\cite{GZ16}. It was extended, still with randomness, to the sum-rank metric~\cite{PRR22}, and rank decoding over finite principal ideal rings is at least as hard as over fields~\cite{KK21}. On the algorithmic side, the best known attacks on rank syndrome decoding take exponential time~\cite{BBB+20,BBC+20}.



	\subsection*{Organization}
	\label{sec:intro:org}

	Section~\ref{sec:prelims} fixes notation and the problem definitions. Section~\ref{sec:lift} proves the explicit diagonal lift and uses repetition to obtain the required extension degree. Section~\ref{body:calibration-section} gives the rank-metric exact-rate calibration. Section~\ref{sec:body-source} derives the quantitative Hamming source from the construction of~\cite{BGR25}. Section~\ref{body:hardness-section} combines these ingredients to prove the rank minimum-distance and decoding results.

	\paragraph*{AI disclosure.}  No part of the research involved any use of AI-based tools.  A complete version of the manuscript was passed through an AI-based tool for recommendations on improving structure and presentation.  Some of these were found useful, and were implemented.  

	\section{Preliminaries}
	\label{sec:prelims}

	Throughout, \(q\) is a fixed prime power and \(\Fq\) is the field with \(q\) elements. For an integer \(m\ge1\), \(\F_{q^m}\) is the field with \(q^m\) elements. All logarithms are base~\(2\), \([N]\coloneqq\{1,\dots,N\}\), and \(\langle S\rangle_K\) denotes the \(K\)-linear span of a set \(S\).

	\subsection*{Linear codes}
	\label{sec:prelims:codes}

	Let \(K\) be a field and \(\Omega\) a finite-dimensional \(K\)-vector space, the \emph{ambient space}. A \emph{(\(K\)-linear) code} is a subspace \(C\le\Omega\). Its \emph{rate} is \(\rate(C)\coloneqq\dim_K(C)/\dim_K(\Omega)\). We use \((K,\Omega)\in\{(\Fq,\Fq^N),(\F_{q^m},\F_{q^m}^N)\}\). Thus a vector code of length \(N\) and dimension \(k\) has rate \(k/N\).


	\subsection*{Extension field.}
	In the rank metric, \(m\) is part of the input. We present \(\F_{q^m}=\Fq[z]/(f(z))\), where \(f(z)\in\Fq[z]\) is monic irreducible and \(\deg_z(f(z))=m\). Each element is stored as a polynomial of degree less than \(m\), using \(O(m\cdot\log q)\) bits. A generator matrix for a code of length \(N\) and dimension \(k\) therefore uses \(\Theta(kNm\cdot\log q)\) bits. We require \(m\le\poly(N)\) throughout.

	An element \(\vartheta\in\F_{q^m}\) has \emph{degree \(d\)} over \(\Fq\) if its minimal polynomial has degree \(d\), equivalently if \(\Fq(\vartheta)=\F_{q^d}\). It has degree exactly \(m\) if and only if \(\Fq(\vartheta)=\F_{q^m}\). Then \(0\ne p(z)\in\Fq[z]\) and \(\deg_z(p(z))<m\) imply \(p(\vartheta)\ne0\). We use this nonvanishing property, together with \(\vartheta\ne0\), for the lift in Section~\ref{sec:lift}.

	\begin{lemma}[Deterministic finite-field construction]
		\label{lem:fields}
		Fix a prime power \(q\). For every \(m\ge1\), one can construct, deterministically in time \(\poly(m)\), a monic irreducible \(f(z)\in\Fq[z]\) of degree \(m\)~\cite{Sho90}. Then \(\F_{q^m}=\Fq[z]/(f(z))\), and \(\vartheta=z\bmod f(z)\) has degree exactly \(m\).
	\end{lemma}

	\begin{proof}
		The construction is given by~\cite{Sho90}. Since \(f(z)\) is the minimal polynomial of \(z\bmod f(z)\), this element has degree \(m\).
	\end{proof}

	\subsection*{Metrics and Operations}
	\label{sec:prelims:metrics}
	\label{sec:prelims:ops}
	\label{sec:prelims:transfers}

	A \emph{metric setting} \(\mathfrak M=(K,\Omega,\wt)\) consists of a field \(K\supseteq\Fq\), a finite-dimensional \(K\)-space \(\Omega\), and a weight \(\wt:\Omega\to\Z_{\ge0}\) that (i) vanishes only at \(0\), (ii) satisfies \(\wt(\lambda x)=\wt(x)\) for \(\lambda\in K^*\), and (iii) satisfies \(\wt(x+y)\le\wt(x)+\wt(y)\).

	For a code \(C\le\Omega\), define \(d(C)\coloneqq\min_{c\in C\setminus\{0\}}\wt(c)\), with \(d(\{0\})\coloneqq+\infty\). For an affine space \(A=a+C\), call \(C\) its \emph{direction code}, and define \(\mu(a+C)\coloneqq\min_{c\in C}\wt(a+c)\) and \(\dist(b,C)\coloneqq\mu(b+C)\). In the rank metric we write \(\muR\) for \(\mu\). We use the following two settings.

	\begin{center}
		\small
		\begin{tabular}{@{}lllll@{}}
			\toprule
			Setting & \(K\) & \(\Omega\) & \(\wt(x)\) & \(d(C)\)\\
			\midrule
			Hamming \(\mathfrak H_N\) & \(\Fq\) & \(\Fq^N\) & \(\|x\|_0=|\{i:x_i\ne0\}|\) & \(\dH(C)\)\\
			Rank \(\mathfrak R_{N,m}\) & \(\F_{q^m}\) & \(\F_{q^m}^N\) & \(\wR(x)=\dim_{\Fq}(\langle x_1,\dots,x_N\rangle_{\Fq})\) & \(\dR(C)\)\\
			\bottomrule
		\end{tabular}
	\end{center}

	\noindent The \emph{relative distance} is \(\delta(C)\coloneqq d(C)/N\). In the rank metric the scalars are \(\F_{q^m}\).


	For \(C\le\Fq^n\) and an integer \(L\ge1\), write \(C^{(L)}\le\Fq^{nL}\) for its \emph{\(L\)-fold coordinate repetition}, the image under
	\[
	\Fq^n\longrightarrow\Fq^{nL},\quad(c_1,\ldots,c_n)\longmapsto(\underbrace{c_1,\ldots,c_1}_{L},\ldots,\underbrace{c_n,\ldots,c_n}_{L}).
	\]
	For \(U\le\Fq^a\) and \(V\le\Fq^b\), define the tensor code \(U\otimes V\coloneqq\{X\in\Fq^{a\times b}:X_{*,j}\in U\ \forall j\in[b],\ X_{i,*}\in V\ \forall i\in[a]\},\) viewed row by row as a code of length \(ab\). For \(C\le\Fq^N\), its scalar extension is \(C_{\F_{q^m}}\coloneqq C\otimes_{\Fq}\F_{q^m}=\langle C\rangle_{\F_{q^m}}\le\F_{q^m}^N\), with the same generator matrix and \(\dim_{\F_{q^m}}(C_{\F_{q^m}})=\dim_{\Fq}(C)\). Direct sums \(C_1\oplus C_2\) concatenate vector blocks; \(0^z\) denotes \(z\) zero coordinates. Lemma~\ref{lem:elementary} records the elementary identities for these operations.

	For \(\alpha\in(\F_{q^m}^*)^N\), let \(D_\alpha\coloneqq\diag(\alpha)\), acting by \(D_\alpha:\F_{q^m}^N\to\F_{q^m}^N,\ x\mapsto\alpha\circ x\). The \emph{diagonal lift}~\cite{GZ16} is \(D_\alpha(C)_{\F_{q^m}}\coloneqq D_\alpha\cdot C_{\F_{q^m}}=\langle\alpha\circ c:c\in C\rangle_{\F_{q^m}}\), with \(\dim_{\F_{q^m}}(D_\alpha(C)_{\F_{q^m}})=\dim_{\Fq}(C)\). If \(C\ne0\), choose \(0\ne c\in C\) with \(\|c\|_0=\dH(C)\). Since \(\alpha\circ c\) has the same support as \(c\), Lemma~\ref{lem:elementary}(2) gives
	\[
	\dR(D_\alpha(C)_{\F_{q^m}})\le\wR(\alpha\circ c)\le\|c\|_0=\dH(C).
	\]
	The inequality can be strict: if \(C\ne0\) and \(\alpha\in(\Fq^*)^N\), the lifted code contains nonzero words with all entries in \(\Fq\), which have rank weight~\(1\). Theorem~\ref{thm:intro-lift} gives an explicit \(\alpha\) achieving equality.

	\subsection*{Problems}
	\label{sec:prelims:promise}\label{sec:prelims:homog}\label{sec:prelims:affine}
	\label{sec:prelims:promises}

	All problems are promise problems \(\Pi=(\Pi_{\YES},\Pi_{\NO})\). An algorithm solves \(\Pi\) if it accepts \(\Pi_{\YES}\) and rejects \(\Pi_{\NO}\), with no requirement elsewhere. A \emph{reduction} is a deterministic polynomial-time map sending \YES{} instances to \YES{} instances and \NO{} instances to \NO{} instances. The parameters \(q,R,\Gamma,\eps\), and the constants used to construct auxiliary codes, are fixed before the input is read. Since a \(\Gamma\)-gap problem reduces to the \(\Gamma'\)-gap problem for every \(\Gamma'<\Gamma\), we may assume \(\Gamma\in\mathbb Q\).

	\begin{definition}[Gap problems]
		\label{def:gapmd}\label{def:gapncp}
		Let \(\mathfrak M=(K,\Omega,\wt)\) be a metric setting, \(\Gamma\ge1\) and \(T\in\Z_{>0}\). In each problem below the \YES{} instances have \(\tx{\normalfont opt}\le T\) and the \NO{} instances have \(\tx{\normalfont opt}>\Gamma T\).
		\begin{itemize}
			\item \emph{Minimum distance} \(\Gamma\)-\texttt{GapMD}\([\mathfrak M]\): given a nonzero code \(C\le\Omega\), \(\tx{\normalfont opt}\coloneqq d(C)\).
			\item \emph{Nearest codeword} \(\Gamma\)-\texttt{GapNC}\([\mathfrak M]\): given a code \(C\le\Omega\) and \(b\in\Omega\), \(\tx{\normalfont opt}\coloneqq\dist(b,C)\). In the \emph{syndrome} form \(\Gamma\)-\texttt{GapSD}\([\mathfrak M]\), one is given a \(K\)-linear map \(H:\Omega\to K^r\) for an integer \(r\ge0\), and a syndrome \(s\in K^r\), with \(\tx{\normalfont opt}\coloneqq\min\{\wt(x):x\in\Omega,\ H(x)=s\}\), which is \(+\infty\) if no solution exists.
		\end{itemize}
		The \emph{rate} of an instance is the rate of its code, or of its direction code \(C=\ker H\).
	\end{definition}

	\noindent Codes are given by a basis. In the rank metric the input also contains \(m\le\poly(N)\) and an irreducible \(f(z)\in\Fq[z]\) with \(\deg_z(f(z))=m\), presenting \(\F_{q^m}=\Fq[z]/(f(z))\) as above.

	\begin{lemma}[Nearest codeword \(\equiv\) consistent syndrome decoding]
		\label{lem:ncp-sd}
		In each setting, \texttt{GapNC} and the restriction of \texttt{GapSD} to consistent systems reduce to each other. The reductions preserve \(\tx{\normalfont opt}\), \(T\), the direction code and its rate. In particular, this applies when \(H\) has full row rank.
	\end{lemma}

	\begin{proof}[Proof of \Cref{lem:ncp-sd}]
		Given \((C,b,T)\), compute a full-row-rank \(H\) with \(\ker H=C\). The reduction and its solution set are
		\[
		(C,b,T)\longmapsto(H,H(b),T),\quad\{x:H(x)=H(b)\}=b+C.
		\]
		Conversely, solve the consistent system \(H(a)=s\), and use
		\[
		(H,s,T)\longmapsto(\ker H,a,T),\quad\{x:H(x)=s\}=a+\ker H.
		\]
		Gaussian elimination detects an inconsistent syndrome system, whose optimum is \(+\infty\). Every nearest-codeword instance has a finite optimum. For hardness of unrestricted syndrome decoding, the consistent outputs already suffice.
	\end{proof}

	\subsection*{Elementary operations on codes}
	\label{subsec:operations}\label{app:routine-proofs}

	\begin{lemma}[Elementary facts]\label{lem:elementary}
		\begin{enumerate}
			\item \emph{(Expansion.)} Fix an \(\Fq\)-basis of \(\F_{q^m}\), and define \(\Phi:\F_{q^m}^N\to\Fq^{m\times N}\) by taking the \(i\)-th column of \(\Phi(x)\) to be the coordinate vector of \(x_i\). Then \(\Phi\) is an \(\Fq\)-linear bijection with \(\wR(x)=\rank\Phi(x)\). For an \(\F_{q^m}\)-linear code \(C\), we have
			\[
			\dim_{\Fq}(\Phi(C))=m\cdot\dim_{\F_{q^m}}(C).
			\]
			\item \emph{(Rank versus Hamming.)} \(\wR(x)\le\min\{\|x\|_0,m\}\) for every \(x\in\F_{q^m}^N\).
			\item \emph{(Direct sums.)} In both settings, \(d(C_1\oplus C_2)=\min\{d(C_1),d(C_2)\}\), \(d(C\oplus0^z)=d(C)\), and \(\mu((a+C)\oplus D)=\mu(a+C)\) for every linear \(D\).
			\item \emph{(Repetition.)} \(C^{(L)}\) has dimension \(\dim(C)\) and \(\dH(C^{(L)})=L\cdot\dH(C)\).
			\item \emph{(Tensoring.)} \(\dim(U\otimes V)=\dim(U)\cdot\dim(V)\) and \(\dH(U\otimes V)=\dH(U)\cdot\dH(V)\).
		\end{enumerate}
	\end{lemma}

	Items~(1) and~(2) are proved below, Item~(3) in the direct-sum lemmas below, and Items~(4) and~(5) in \Cref{lem:repetition-full,lem:tensor-full}.

	\label{body:expansion-proof}
	The nonzero coordinates of \(x\) generate a subspace of the \(m\)-dimensional \(\F_q\)-space \(\F_{q^m}\), so
	\[
	\wR(x)\le\min\{\|x\|_0,m\}\le\min\{N,m\},\quad x\in \F_{q^m}^N.
	\]
	In particular, \(\dR(\cC)\le\min\{N,m\}\) for every nonzero \(\cC\le \F_{q^m}^N\). Rank weight is at most Hamming weight, and can be much smaller.

	Fix an \(\F_q\)-basis \(\xi=(\xi_1,\dots,\xi_m)\) of \(\F_{q^m}\), and write \([a]_\xi\in\F_q^m\) for the coordinate column of \(a\in \F_{q^m}\). Coordinate expansion is the \(\F_q\)-linear isomorphism
	\[
	M:\F_{q^m}^N\longrightarrow\F_q^{m\times N},\quad x\longmapsto M(x)\coloneqq\bigl[[x_1]_\xi\ \cdots\ [x_N]_\xi\bigr].
	\]
	The coordinate isomorphism \(a\mapsto[a]_\xi\) gives
	\[
	\operatorname{col}M(x)=\langle[x_1]_\xi,\dots,[x_N]_\xi\rangle_{\F_q}\cong\langle x_1,\dots,x_N\rangle_{\F_q},\quad\rank_{\F_q}M(x)=\wR(x).
	\]
	For an \(\F_{q^m}\)-linear code \(\cC\le \F_{q^m}^N\), its restriction gives
	\[
	\dim_{\F_q}(M(\cC))=\dim_{\F_q}(\cC)=m\cdot\dim_{\F_{q^m}}(\cC).
	\]
	Lemma~\ref{lem:rat} expresses the same weight identity using the rational subspace \(\Rat(x)\).

	If the rows \(c_1,\dots,c_k\) of \(G\in\F_q^{k\times N}\) form an \(\F_q\)-basis of \(C\), then
	\[
	\F_{q^m}^k\longrightarrow C_{\F_{q^m}},\quad(\lambda_1,\dots,\lambda_k)\longmapsto\sum_{i=1}^k\lambda_ic_i=\lambda G
	\]
	is an isomorphism: it is surjective by definition, and a nonzero \(k\times k\) minor of \(G\) over \(\F_q\) remains nonzero over \(\F_{q^m}\). Thus \(\dim_{\F_{q^m}}(C_{\F_{q^m}})=k=\dim_{\F_q}(C)\).

	For codes \(C_1\le\F_q^N\) and \(C_2\le\F_q^{N'}\), their direct sum is
	\[
	C_1\oplus C_2\coloneqq\{(c_1,c_2):c_1\in C_1,\ c_2\in C_2\}\le\F_q^{N+N'},
	\]
	and the same definition applies over \(\F_{q^m}\). We write \(0^z\) for the zero-dimensional code of length \(z\), in either metric. The next two lemmas give the dimension and minimum distance of direct sums in the two metrics.

	\begin{lemma}[Direct sums and zero padding, Hamming metric]
		\label{lem:hamming-directsum}
		Let \(C_1\le\F_q^N\) and \(C_2\le\F_q^{N'}\) be nonzero linear codes. Then
		\[
		\dim_{\F_q}(C_1\oplus C_2)=\dim_{\F_q}(C_1)+\dim_{\F_q}(C_2), \quad \dH(C_1\oplus C_2)=\min\{\dH(C_1),\dH(C_2)\}.
		\]
		For a nonzero code \(C\le\F_q^N\) and an integer \(z\ge0\), \(\dim_{\F_q}(C\oplus0^z)=\dim_{\F_q}(C)\) and \(\dH(C\oplus0^z)=\dH(C)\).
	\end{lemma}

	\begin{proof}
		Let \(b_1,\dots,b_k\) and \(b'_1,\dots,b'_{k'}\) be bases of \(C_1\) and \(C_2\). The vectors \((b_i,0)\) and \((0,b'_j)\) span \(C_1\oplus C_2\), since
		\[
		\left(\sum_i\lambda_ib_i,\sum_j\mu_jb'_j\right)=\sum_i\lambda_i(b_i,0)+\sum_j\mu_j(0,b'_j).
		\]
		They are independent, since
		\[
		\begin{aligned}
			\sum_i\lambda_i(b_i,0)+\sum_j\mu_j(0,b'_j)=0&\iff\sum_i\lambda_ib_i=0\ \tx{and}\ \sum_j\mu_jb'_j=0,\\
			&\iff\lambda_i=\mu_j=0\ \tx{for all }i,j.
		\end{aligned}
		\]
		Thus \(\dim_{\F_q}(C_1\oplus C_2)=k+k'\).

		For any nonzero \((c_1,c_2)\in C_1\oplus C_2\), the disjoint coordinate blocks give
		\[
		\|(c_1,c_2)\|_0=\|c_1\|_0+\|c_2\|_0\ge\min\{\dH(C_1),\dH(C_2)\}.
		\]
		Choose minimum-weight nonzero \(c_1\in C_1\) and \(c_2\in C_2\). The reverse bound follows from
		\[
		\|(c_1,0)\|_0=\dH(C_1),\quad\|(0,c_2)\|_0=\dH(C_2).
		\]
		Hence \(\dH(C_1\oplus C_2)=\min\{\dH(C_1),\dH(C_2)\}\). Finally, the isomorphism \(C\to C\oplus0^z\), \(c\mapsto(c,0^z)\) satisfies \(\|(c,0^z)\|_0=\|c\|_0\), proving both zero-padding assertions.
	\end{proof}

	\begin{lemma}[Direct sums and zero padding, rank metric]
		\label{lem:rank-directsum}
		Let \(\cC_1\le \F_{q^m}^{N_1}\) and \(\cC_2\le \F_{q^m}^{N_2}\) be nonzero \(\F_{q^m}\)-linear rank-metric codes. Then
		\[
		\dim_{\F_{q^m}}(\cC_1\oplus\cC_2)=\dim_{\F_{q^m}}(\cC_1)+\dim_{\F_{q^m}}(\cC_2), \quad \dR(\cC_1\oplus\cC_2)=\min\{\dR(\cC_1),\dR(\cC_2)\}.
		\]
		For a nonzero code \(\cC\le \F_{q^m}^N\) and an integer \(z\ge0\), \(\dim_{\F_{q^m}}(\cC\oplus0^z)=\dim_{\F_{q^m}}(\cC)\) and \(\dR(\cC\oplus0^z)=\dR(\cC)\).
	\end{lemma}

	\begin{proof}
		The dimension follows from the basis argument in \Cref{lem:hamming-directsum}, now over \(\F_{q^m}\). For \((x,y)\in\cC_1\oplus\cC_2\), the coordinate spans give
		\[
		\wR(x,y)=\dim_{\F_q}\bigl(\langle x_1,\dots,x_{N_1}\rangle_{\F_q}+\langle y_1,\dots,y_{N_2}\rangle_{\F_q}\bigr)\ge\max\{\wR(x),\wR(y)\}.
		\]
		If \((x,y)\ne0\), at least one component is nonzero, so
		\[
		\wR(x,y)\ge\min\{\dR(\cC_1),\dR(\cC_2)\}.
		\]
		For minimum-weight nonzero \(x\in\cC_1\) and \(y\in\cC_2\),
		\[
		\wR(x,0)=\wR(x)=\dR(\cC_1),\quad\wR(0,y)=\wR(y)=\dR(\cC_2).
		\]
		These two codewords give the reverse inequality. For zero padding, the map \(\cC\to\cC\oplus0^z,\ x\mapsto(x,0^z)\) is an \(\F_{q^m}\)-linear isomorphism, and
		\[
		\langle x_1,\dots,x_N,0,\dots,0\rangle_{\F_q}=\langle x_1,\dots,x_N\rangle_{\F_q}.
		\]
		Thus it preserves dimension and rank distance.
	\end{proof}

	To set the rate later, we will also use the following fact about affine spaces in either metric. If \(A=a+C\) is an affine space and \(D\) is a linear code in the same metric, then adjoining a new coordinate block ranging over \(D\) does not change the minimum weight:
	\[
	\mu(A\oplus D)=\mu(A).
	\]
	Write \((x,d)\) for the concatenated vector. The preceding identities give
	\[
	\wgt(x,d)\ge\wgt(x),\quad\wgt(x,0)=\wgt(x),\quad x\in A,\ d\in D,
	\]
	and therefore
	\[
	\min_{x\in A,\,d\in D}\wgt(x,d)=\min_{x\in A}\wgt(x)=\mu(A).
	\]
	Appending zero coordinates preserves every weight, so it preserves both the minimum weight in an affine space and the minimum nonzero weight of a code. We use these facts with \Cref{lem:hamming-directsum,lem:rank-directsum}.

	\begin{lemma}[Tensor-product distance]
		\label{lem:tensor-full}
		For nonzero linear codes \(U\le\F_q^{a}\) and \(V\le\F_q^{b}\), let \(U\otimes V\) be the space of matrices \(M\in\F_q^{a\times b}\) whose columns lie in \(U\) and whose rows lie in \(V\). Then
		\[
		\dH(U\otimes V)=\dH(U)\cdot\dH(V),
		\]
		and consequently, for every integer \(s\ge1\), \(\dH(U^{\otimes s})=\dH(U)^s\).
	\end{lemma}

	\begin{proof}
		We first give a basis of this matrix space. Let \(u_1,\dots,u_k\) and \(v_1,\dots,v_h\) be bases of \(U\) and \(V\), and put \(P\coloneqq[u_1\ \cdots\ u_k]\in\F_q^{a\times k}\) and \(Q\coloneqq[v_1\ \cdots\ v_h]\in\F_q^{b\times h}\). Choose left inverses \(L\in\F_q^{k\times a}\) and \(J\in\F_q^{h\times b}\), so \(LP=I_k\) and \(JQ=I_h\). If \(M\in U\otimes V\), its columns lie in \(U\), so \(M=P(LM)\). Each row of \(LM\) is a linear combination of rows of \(M\) and therefore lies in \(V\). Thus
		\[
		M=PAQ^{\transpose}=\sum_{i=1}^k\sum_{j=1}^h A_{ij}u_iv_j^{\transpose}\quad\tx{for some }A\in\F_q^{k\times h}.
		\]
		Conversely, every such matrix has columns in \(U\) and rows in \(V\). Further,
		\[
		PAQ^{\transpose}=0\implies L(PAQ^{\transpose})J^{\transpose}=A=0.
		\]
		Hence the matrices \(u_iv_j^{\transpose}\) form a basis and \(\dim_{\F_q}(U\otimes V)=kh\).

		For the upper bound, choose minimum-weight nonzero \(u\in U\) and \(v\in V\). Then \(uv^{\transpose}\in U\otimes V\), and
		\[
		\supp(uv^{\transpose})=\supp(u)\times\supp(v),\quad\|uv^{\transpose}\|_0=\|u\|_0\|v\|_0=\dH(U)\cdot\dH(V).
		\]
		For the lower bound, let \(0\ne M\in U\otimes V\), choose a nonzero column \(M_{*,j}\), and set \(S\coloneqq\{i:M_{i,j}\ne0\}\). Since \(M_{*,j}\in U\setminus\{0\}\) and \(M_{i,*}\in V\setminus\{0\}\) for \(i\in S\),
		\[
		|S|\ge\dH(U),\quad\|M\|_0=\sum_i\|M_{i,*}\|_0\ge\sum_{i\in S}\|M_{i,*}\|_0\ge|S|\cdot\dH(V)\ge\dH(U)\cdot\dH(V).
		\]
		This proves the two-factor distance identity. The case \(s=1\) is immediate, and induction gives, for \(s\ge2\),
		\[
		\dH(U^{\otimes s})=\dH(U)\cdot\dH(U^{\otimes(s-1)})=\dH(U)\cdot\dH(U)^{s-1}=\dH(U)^s.
		\]
		The same basis calculation gives \(\dim_{\F_q}(U^{\otimes s})=(\dim_{\F_q}(U))^s\).
	\end{proof}

	\begin{lemma}[Coordinate repetition]
		\label{lem:repetition-full}
		Let \(C\le\F_q^{n_0}\) be a nonzero code of dimension \(k\), and let \(L\ge1\) be an integer. Let
		\[
		C^{(L)}\coloneqq\Bigl\{\bigl(\underbrace{c_1,\dots,c_1}_{L},\underbrace{c_2,\dots,c_2}_{L},\dots,\underbrace{c_{n_0},\dots,c_{n_0}}_{L}\bigr):c\in C\Bigr\}\le\F_q^{n_0L}
		\]
		be its \(L\)-fold coordinate repetition, in which each coordinate is repeated \(L\) consecutive times. Then \(C^{(L)}\) has length \(n_0L\), dimension \(k\), and
		\[
		\dH(C^{(L)})=L\cdot\dH(C).
		\]
		Moreover, for every threshold \(t\) and every \(\gamma\ge1\), the map \(C\mapsto C^{(L)}\) with threshold \(Lt\) preserves the YES and NO alternatives for Hamming minimum distance at gap \(\gamma\).
	\end{lemma}

	\begin{proof}
		The \(\F_q\)-linear repetition map and its left inverse are
		\[
		C\longrightarrow C^{(L)},\quad c\longmapsto c^{(L)},\quad C^{(L)}\longrightarrow C,\quad y\longmapsto(y_{(i-1)L+1})_{i=1}^{n_0}.
		\]
		They are inverse isomorphisms, so \(\dim_{\F_q}(C^{(L)})=\dim_{\F_q}(C)=k\). Writing \([\cdot]\) for an indicator,
		\[
		\|c^{(L)}\|_0=\sum_{i=1}^{n_0}L[c_i\ne0]=L\sum_{i=1}^{n_0}[c_i\ne0]=L\|c\|_0.
		\]
		Taking the minimum over nonzero \(c\in C\) gives \(\dH(C^{(L)})=L\cdot\dH(C)\). Thus
		\[
		\dH(C)\le t\implies\dH(C^{(L)})\le Lt,\quad\dH(C)>\gamma t\implies\dH(C^{(L)})>\gamma Lt.
		\]
	\end{proof}


	\section[A deterministic distance-preserving lift into the rank metric]{A deterministic distance-preserving lift into the rank metric\\
		{\small Proof of Theorem~\ref{thm:intro-lift}.}}
	\label{sec:lift}

	Let \(C_0\le\F_q^N\) be a nonzero code of dimension \(k\), and put \(d_0\coloneqq\dH(C_0)\). The upper bound \(\dR(D_\alpha(C_0)_{\F_{q^m}})\le d_0\) holds for every multiplier, so we need the reverse inequality for \(\alpha=(1,\vartheta,\dots,\vartheta^{N-1})\). The difficulty is that linear combinations over the extension field can have small rank. We will rule this out by showing that the lifted code has zero intersection with certain subspaces defined over \(\F_q\). We first prove this over \(\F_q(t)\), bound the spread of a nonzero minor, and then substitute \(\vartheta\) for \(t\).

	\subsection{Rational subspaces}
	\label{subsec:rat}

	Fix a finite extension \(\F_{q^m}/\F_q\) and an \(\F_q\)-basis \(\xi_1,\dots,\xi_m\) of \(\F_{q^m}\). Every \(y\in \F_{q^m}^N\) has a unique expansion, and we define its rational subspace by
	\[
	y=\sum_{a=1}^{m}\xi_a y^{(a)},\quad y^{(a)}\in\F_q^N,\quad\Rat(y)\coloneqq\langle y^{(1)},\dots,y^{(m)}\rangle_{\F_q}.
	\]

	\begin{lemma}[Rational subspaces]
		\label{lem:rat}
		For every \(y\in \F_{q^m}^N\), the subspace \(\Rat(y)\) is independent of the chosen basis of \(\F_{q^m}\). It is the smallest \(\F_q\)-subspace \(U\le\F_q^N\) with \(y\in U_{\F_{q^m}}\), and \(\dim_{\F_q}(\Rat(y))=\wR(y)\). Consequently, for a nonzero code \(C_0\le\F_q^N\) with \(\alpha\in(\F_{q^m}^\ast)^N\) and an integer \(d\), one has \(\wR(y)<d\) for some nonzero \(y\in D_\alpha(C_0)_{\F_{q^m}}\) if and only if \(D_\alpha(C_0)_{\F_{q^m}}\cap U_{\F_{q^m}}\ne\{0\}\) for some subspace \(U\le\F_q^N\) with \(\dim_{\F_q}(U)<d\).
	\end{lemma}

	\begin{proof}[Proof of \Cref{lem:rat}]
		Form the matrix \((y_i^{(a)})_{a\in[m],i\in[N]}\in\F_q^{m\times N}\), whose \(a\)-th row is \(y^{(a)}\). The coordinate isomorphism
		\[
		\F_{q^m}\longrightarrow\F_q^m,\quad \sum_{a=1}^m\xi_ac_a\longmapsto(c_1,\ldots,c_m)^{\transpose}
		\]
		sends \(y_i\) to its \(i\)-th column. Thus its row space is \(\Rat(y)\), and its column space is isomorphic to \(\langle y_1,\ldots,y_N\rangle_{\F_q}\). Equality of row and column rank gives
		\begin{align*}
			\dim_{\F_q}(\Rat(y))&=\rank_{\F_q}(y_i^{(a)})_{a,i}\\
			&=\dim_{\F_q}(\langle y_1,\ldots,y_N\rangle_{\F_q})=\wR(y).
		\end{align*}

		Fix a subspace \(U\le\F_q^N\) and an \(\F_q\)-basis \(u_1,\ldots,u_r\) of \(U\). This is also an \(\F_{q^m}\)-basis of \(U_{\F_{q^m}}\). If \(y\in U_{\F_{q^m}}\), write
		\[
		y=\sum_{b=1}^re_bu_b,\quad e_b=\sum_{a=1}^m\xi_ae_b^{(a)},\quad e_b^{(a)}\in\F_q.
		\]
		By uniqueness of the expansion in the basis \(\xi_1,\ldots,\xi_m\),
		\[
		y=\sum_{a=1}^m\xi_a\left(\sum_{b=1}^re_b^{(a)}u_b\right)\quad\Longrightarrow\quad y^{(a)}=\sum_{b=1}^re_b^{(a)}u_b\in U\quad\tx{for every }a\in[m].
		\]
		Conversely, if \(y^{(a)}\in U\) for every \(a\), then \(y=\sum_a\xi_ay^{(a)}\in U_{\F_{q^m}}\). Therefore
		\[
		y\in U_{\F_{q^m}}\iff y^{(a)}\in U\ \tx{for every }a\in[m]\iff\Rat(y)\le U.
		\]
		Hence \(\Rat(y)\) is the smallest \(\F_q\)-subspace \(U\) with \(y\in U_{\F_{q^m}}\). This characterization does not depend on the chosen basis.

		For \(0\ne y\in D_\alpha(C_0)_{\F_{q^m}}\), the identities above give
		\begin{align*}
			\wR(y)<d&\quad\Longrightarrow\quad U\coloneqq\Rat(y),\quad\dim_{\F_q}(U)<d,\quad y\in U_{\F_{q^m}},\\
			y\in U_{\F_{q^m}},\quad\dim_{\F_q}(U)<d&\quad\Longrightarrow\quad\wR(y)=\dim_{\F_q}(\Rat(y))\le\dim_{\F_q}(U)<d.
		\end{align*}
		Thus a nonzero word of rank weight less than \(d\) exists in \(D_\alpha(C_0)_{\F_{q^m}}\) if and only if \(D_\alpha(C_0)_{\F_{q^m}}\cap U_{\F_{q^m}}\ne\{0\}\) for some \(U\le\F_q^N\) with \(\dim_{\F_q}(U)<d\), as required.
	\end{proof}

	By \Cref{lem:rat}, the required lower bound is equivalent to
	\[
	D_\alpha(C_0)_{\F_{q^m}}\cap U_{\F_{q^m}}=\{0\}\quad\tx{for every }U\le\F_q^N\tx{ with }\dim_{\F_q}(U)<d_0.
	\]

	\subsection{An intersection criterion}
	\label{subsec:intersection}

	We express the required zero intersection as a rank condition, or equivalently as the nonvanishing of a minor. This criterion applies both over \(\F_q(t)\) and over \(\F_{q^m}\). Let \(G\in\F_q^{N\times k}\) and \(B\in\F_q^{N\times r}\) have columns forming bases of \(C_0\) and \(U\), respectively, where \(r\coloneqq\dim_{\F_q}(U)\). If \(r=0\), then \(B\) is empty.

	\begin{lemma}[Intersection criterion]
		\label{lem:intersection}
		Let \(F\) be a field containing \(\F_q\) and let \(D\in F^{N\times N}\) be an invertible diagonal matrix. Then, with \(C_0\) and \(U\) as above, \(D\,(C_0)_F\cap U_F=\{0\}\iff\rank_F[\,DG\mid B\,]=k+r,\) and this rank is full if and only if some \((k+r)\times(k+r)\) minor of \([\,DG\mid B\,]\) is nonzero.
	\end{lemma}

	By the Singleton bound, \(r<d_0\le N-k+1\), so \(r\le d_0-1\le N-k\) and \(k+r\le N\).

	\begin{proof}[Proof of \Cref{lem:intersection}]
		Over \(F\), the spaces \(D(C_0)_F\) and \(U_F\) are the column spans of \(DG\in F^{N\times k}\) and \(B\in F^{N\times r}\), respectively. Since \(D\) is invertible and the columns of \(G\) and \(B\) are independent, we have \(\ker(DG)=\{0\}\) and \(\ker B=\{0\}\).

		If \(0\ne v\in D(C_0)_F\cap U_F\), then
		\[
		v=DGz=Bw,\quad z\in F^k\setminus\{0\},\quad w\in F^r,\quad [\,DG\mid B\,]\binom{z}{-w}=0.
		\]
		This is a nontrivial dependence among the columns. Conversely, a nontrivial dependence has \(DGz=Bw\) with \((z,w)\ne0\). Then
		\[
		z=0\Longrightarrow Bw=0\Longrightarrow w=0,
		\]
		a contradiction. Thus \(z\ne0\), and \(\ker(DG)=\{0\}\) gives \(0\ne DGz=Bw\in D(C_0)_F\cap U_F\).

		Since \(N\ge k+r\), we conclude
		\begin{align*}
			D(C_0)_F\cap U_F=\{0\}&\iff\rank_F[\,DG\mid B\,]=k+r\\
			&\iff\exists R\subseteq[N],\ |R|=k+r:\ \det([\,DG\mid B\,]_R)\ne0.
		\end{align*}
	\end{proof}

	\subsection{Intersection over the function field}
	\label{subsec:symbolic}

	We first prove that the intersection is zero over \(K\coloneqq\F_q(t)\), with \(D(t)\coloneqq\diag(1,t,t^2,\dots,t^{N-1})\).

	\begin{lemma}[Intersection over the function field]
		\label{lem:symbolic}
		Let \(U\le\F_q^N\) have dimension \(r<d_0\). Then \(D(t)\,(C_0)_K\cap U_K=\{0\}\), equivalently \(\rank_K[\,D(t)G\mid B\,]=k+r\).
	\end{lemma}

	\begin{proof}
		Suppose that \(0\ne y(t)\in D(t)(C_0)_K\cap U_K\), and write \(y(t)=D(t)x(t)\), where \(0\ne x(t)\in(C_0)_K\). Fix \(\F_q\)-bases \(c^{(1)},\dots,c^{(k)}\) of \(C_0\) and \(u^{(1)},\dots,u^{(r)}\) of \(U\), and write
		\[
		x(t)=\sum_{b=1}^{k}\lambda_b(t)c^{(b)},\quad y(t)=\sum_{b'=1}^{r}\mu_{b'}(t)u^{(b')},\quad\lambda_b(t),\mu_{b'}(t)\in K.
		\]
		Choose \(0\ne p(t)\in\F_q[t]\) with \(p(t)\lambda_b(t),p(t)\mu_{b'}(t)\in\F_q[t]\) for all indices. Multiplying by \(p(t)\) keeps both vectors nonzero and in their respective spaces, and \(p(t)y(t)=D(t)(p(t)x(t))\). Replace \(x(t),y(t),\lambda_b(t),\mu_{b'}(t)\) by their multiples by \(p(t)\), retaining the notation. Then \(x(t)=\sum_s x^{[s]}t^s\) and \(y(t)=\sum_s y^{[s]}t^s\) have finite support, with
		\[
		x^{[s]}=\sum_{b=1}^{k}([t^s]\lambda_b(t))c^{(b)}\in C_0,\quad y^{[s]}=\sum_{b'=1}^{r}([t^s]\mu_{b'}(t))u^{(b')}\in U.
		\]
		Also, \(y_i(t)=t^{i-1}x_i(t)\) for every \(i\in[N]\).

		We first show that the coordinates of \(y(t)\) span a space of dimension at most \(r\) over \(\F_q\). Put \(W\coloneqq\langle y_1(t),\dots,y_N(t)\rangle_{\F_q}\), \(\cS\coloneqq\{s\in\Z:y^{[s]}\ne0\}\), and
		\[
		Y\coloneqq\bigl([t^s]y_i(t)\bigr)_{s\in\cS,\,i\in[N]}\in\F_q^{\cS\times[N]},\quad\Phi:W\longrightarrow\F_q^{\cS},\quad h(t)\longmapsto([t^s]h(t))_{s\in\cS}.
		\]
		The set \(\cS\) is finite, every \(h(t)\in W\) is supported on \(\cS\), and \(\Phi(y_i(t))=Y_{*,i}\). Thus \(\ker\Phi=\{0\}\) and \(\Phi(W)\) is the column space of \(Y\). Since \(Y_{s,*}=y^{[s]}\in U\),
		\[
		\dim_{\F_q}(W)=\dim_{\F_q}(\Phi(W))=\rank_{\F_q}Y\le\dim_{\F_q}(U)=r.
		\]

		Select coordinates from left to right, putting \(T\coloneqq\{j\in[N]:y_j(t)\notin\langle y_1(t),\dots,y_{j-1}(t)\rangle_{\F_q}\}\), where the span of the empty set is \(\{0\}\). At step \(j\), a dependent \(y_j(t)\) leaves the span unchanged; otherwise it extends the basis. Thus, by induction, \(\{y_i(t):i\in T\cap[j]\}\) is a basis of \(\langle y_i(t):i\le j\rangle_{\F_q}\), and \(|T|=\dim_{\F_q}(W)\le r\). For \(j\notin T\), choose \(c_{j,i}\in\F_q\) such that
		\[
		y_j(t)=\sum_{\substack{i\in T\\
				i<j}}c_{j,i}y_i(t)\implies x_j(t)=\sum_{\substack{i\in T\\
				i<j}}c_{j,i}t^{i-j}x_i(t).
		\]

		Define \(\sigma^*\coloneqq\max_{i:x_i(t)\ne0}\deg_t(x_i(t))\). Every nonzero summand on the right has degree \(\deg_t(c_{j,i}t^{i-j}x_i(t))=i-j+\deg_t(x_i(t))\le\sigma^*-1\). Therefore \([t^{\sigma^*}]x_j(t)=0\) for \(j\notin T\). The coefficient vector \(c\coloneqq x^{[\sigma^*]}=([t^{\sigma^*}]x_i(t))_{i\in[N]}\) satisfies
		\[
		0\ne c\in C_0,\quad\supp(c)\subseteq T\implies d_0\le\|c\|_0\le|T|\le r<d_0,
		\]
		a contradiction. Hence the intersection is zero, and \Cref{lem:intersection} gives \(\rank_K[\,D(t)G\mid B\,]=k+r\).
	\end{proof}

	\subsection{Degree spread of the maximal minors}
	\label{subsec:spread}

	By \Cref{lem:symbolic}, for every \(U\) with \(r=\dim_{\F_q}(U)<d_0\) there is a row set \(R\subseteq[N]\), \(|R|=k+r\), for which \(P_R(t)\coloneqq\det([\,D(t)G\mid B\,]_R)\ne0\). For a nonzero Laurent polynomial \(p(t)\), define \(\operatorname{spread}(p(t))\coloneqq\max\{s:[t^s]p(t)\ne0\}-\min\{s:[t^s]p(t)\ne0\}\).

	\begin{lemma}[Spread bound]
		\label{lem:spread}
		Every nonzero maximal minor \(P_R(t)\) of \([\,D(t)G\mid B\,]\) satisfies \(\operatorname{spread}(P_R(t))\le k(N-k)\).
	\end{lemma}
	\begin{proof}
		Write \(R=\{r_1<\cdots<r_{k+r}\}\). Expanding along the first \(k\) columns gives
		\[
		P_R(t)=\sum_{\substack{S\subseteq R\\
				|S|=k}}\eps_{R,S}\,t^{\sum_{i\in S}(i-1)}\cdot\det(G_{S,[k]})\cdot\det(B_{R\setminus S,[r]}),
		\]
		where \(\eps_{R,S}\coloneqq\sgn(k(k+1)/2+\sum_{\ell:r_\ell\in S}\ell)\), and the empty determinant is \(1\). For every such \(S\), \(k(k-1)/2\le\sum_{i\in S}(i-1)\le k(2N-k-1)/2\). Cancellation can only remove terms, so \(\operatorname{spread}(P_R(t))\le k(2N-k-1)/2-k(k-1)/2=k(N-k)\).
	\end{proof}

	\subsection{Proof of~\cref{thm:intro-lift}}
	\label{subsec:specialize}

	We now substitute \(\vartheta\) for \(t\) and prove~\cref{thm:intro-lift}.
	\begin{proof}[Proof of Theorem~\ref{thm:intro-lift}]
		By \Cref{lem:rat}, it remains to show \(D_\alpha(C_0)_{\F_{q^m}}\cap U_{\F_{q^m}}=\{0\}\) for every \(U\le\F_q^N\) with \(r=\dim_{\F_q}(U)<d_0\). Fix such a \(U\), with basis matrix \(B\). By \Cref{lem:symbolic,lem:spread}, some row set \(R\) of size \(k+r\) gives \(P_R(t)\ne0\) with spread at most \(k(N-k)\). Write \(P_R(t)=t^aQ_R(t)\), where \(a\ge0\), \(Q_R(t)\in\F_q[t]\), and \(Q_R(0)\ne0\). Then \(\deg_t(Q_R(t))=\operatorname{spread}(P_R(t))\le k(N-k)<m.\) Since \(\vartheta\) has degree \(m\) over \(\F_q\), we have \(Q_R(\vartheta)\ne0\). Also \(\vartheta\ne0\) and \(D(\vartheta)=D_\alpha\), so
		\[
		0\ne\vartheta^aQ_R(\vartheta)=P_R(\vartheta)=\det([\,D_\alpha G\mid B\,]_R)\implies\rank_{\F_{q^m}}[\,D_\alpha G\mid B\,]=k+r.
		\]
		By \Cref{lem:intersection}, \(D_\alpha(C_0)_{\F_{q^m}}\cap U_{\F_{q^m}}=\{0\}\). Since \(U\) was arbitrary, \Cref{lem:rat} and the upper bound from Section~\ref{sec:prelims} give \(d_0\le\dR(D_\alpha(C_0)_{\F_{q^m}})\le d_0\).
	\end{proof}


	\subsection{Repetition and the extension degree}
	\label{body:repetition-section}

	Repetition lowers the exponent in the extension-degree bound. We also require \(m\ge\kappa N_1\), so that the same field can support the auxiliary code used for rate calibration.

	\begin{lemma}[Repetition before lifting]
		\label{body:repetition}
		Fix \(q\), \(\eps>0\), and a constant \(\kappa\ge1\). Let \(B\coloneqq\max\{1,\lceil1/\eps\rceil\}\) be an integer. Given a nonzero code \(C\le\F_q^{n_0}\) of dimension \(k\), put
		\[
		L\coloneqq n_0^{B-1},\quad N_1\coloneqq n_0^B,\quad m\coloneqq\max\{2,k(N_1-k)+1,\lceil\kappa N_1\rceil\}.
		\]
		One can construct deterministically an \(\F_{q^m}\)-linear code \(\cC\le \F_{q^m}^{N_1}\) such that \(\dim_{\F_{q^m}}(\cC)=k,\quad\dR(\cC)=L\cdot\dH(C),\quad m=O_{\kappa,\eps}(N_1^{1+\eps}).\) The construction is polynomial time for fixed \(q,\kappa,\eps\).
	\end{lemma}

	\begin{proof}
		Repeat each coordinate of \(C\) exactly \(L\) times to obtain \(C^{(L)}\). Construct \(\F_{q^m}\) and an element \(\vartheta\) of degree \(m\) over \(\F_q\) by \Cref{lem:fields}. Our choice of degree gives \(m\ge2\), \(m>k(N_1-k)\), and \(m\ge\kappa N_1\); in particular, \(\vartheta\ne0\). Define \(\alpha\coloneqq(1,\vartheta,\ldots,\vartheta^{N_1-1})\) and \(\cC\coloneqq D_\alpha(C^{(L)})_{\F_{q^m}}\). Since \(D_\alpha\) is invertible, scalar extension and \Cref{thm:intro-lift} give \(\dim_{\F_{q^m}}(\cC)=\dim_{\F_q}(C^{(L)})=k,\quad\dR(\cC)=\dH(C^{(L)})=L\cdot\dH(C).\) Finally,
		\[
		k\le n_0=N_1^{1/B},\quad\frac1B\le\eps\implies m=O_\kappa(N_1^{1+1/B})=O_{\kappa,\eps}(N_1^{1+\eps}).
		\]
		The generator matrix uses \(O(kN_1m\cdot\log q)\) bits, which is bounded by a polynomial in \(n_0\) for fixed \(q,\kappa,\eps\). Coordinate repetition, finite-field construction, and diagonal scaling all take polynomial time.
	\end{proof}

	\section{Exact rate calibration}
	\label{body:calibration-section}

	We now change the rate while preserving minimum distance and the minimum weight in a given affine space. Throughout this section, we work over \(\F_{q^m}\) with rank weight over \(\F_q\).

	\begin{lemma}[Auxiliary codes]
		\label{body:outer}
		Fix \(q\) and a rational \(R\in(0,1)\). For integers \(n\ge1\) and \(0\le k\le n\), one can construct a nonzero code \(D\le \F_{q^m}^\ell\) of dimension \(h\) satisfying \(\dR(D)\ge n\), \(k+h\ge R(n+\ell)\), and \(\ell=O_R(n)\). The constants depend only on \(R\), and the construction requires \(m\ge\kappa_R n\) for a constant \(\kappa_R\ge1\). The construction is deterministic and polynomial time in its input and output sizes.
	\end{lemma}

	The condition \(\dR(D)\ge n\) ensures that adjoining \(D\) does not lower the input distance. The inequality \(k+h\ge R(n+\ell)\) ensures that the number of zero coordinates added below is nonnegative.

	\begin{proof}[Proof of \Cref{body:outer}]
		Put \(\sigma\coloneqq(1+R)/2\) and \(\Delta\coloneqq1-\sigma\), and put
		\[
		M\coloneqq\left\lceil\max\left\{\frac n\Delta,\frac{(Rn-k)_+}{\sigma-R},1\right\}\right\rceil,\quad (x)_+\coloneqq\max\{x,0\}.
		\]
		Set \(\ell\coloneqq M\). For
		\[
		\kappa_R\coloneqq1+\max\{(1-\sigma)^{-1},R/(\sigma-R),1\},
		\]
		we have \(\ell\le\kappa_R n\le m\). Thus \Cref{lem:gabidulin} gives a Gabidulin code \(D\) with \(h\coloneqq\lceil\sigma\ell\rceil\) and
		\[
		\dR(D)=\ell-h+1=\ell-\lceil\sigma\ell\rceil+1\ge(1-\sigma)\ell\ge n.
		\]

		Then \(D\ne0\), \(\ell=O(n)\), and
		\[
		k+h-R(n+\ell)\ge k-Rn+(\sigma-R)\ell\ge0.
		\]
		The construction and its running time follow from~\Cref{lem:gabidulin}.
	\end{proof}

	\begin{lemma}[Rate calibration for linear codes and affine spaces]
		\label{body:calibration}
		Fix \(q\) and a rational \(R=a/b\in(0,1)\) in lowest terms. Given \(C\le \F_{q^m}^n\), where \(n\ge1\), there is a deterministic construction of a nonzero code \(W\le \F_{q^m}^N\) such that
		\[
		\rate(W)=R,\quad n\le N\le cn,\quad
		\begin{cases}
			\dR(W)=\dR(C),&C\ne0,\\
			\dR(W)\ge n,&C=0.
		\end{cases}
		\]
		Given additionally \(u+C\subseteq \F_{q^m}^n\), the same construction produces \(u^\sharp+W\) with \(\muR(u^\sharp+W)=\muR(u+C)\). Here \(c=c_R\), and the construction requires \(m\ge\kappa_Rn\). Thresholds are unchanged. If \(T\ge\theta n\) and either \(C=0\) or \(\dR(C)\ge\eta n\), where \(0<\eta\le1\), then \(T\ge(\theta/c)N\) and \(\dR(W)\ge(\eta/c)N\).
	\end{lemma}

	\begin{proof}
		Set \(k\coloneqq\dim_{\F_{q^m}}(C)\), and choose \(D\) by \Cref{body:outer}. Put \(U\coloneqq C\oplus D\), \(n_U\coloneqq n+\ell\), \(k_U\coloneqq k+h\), \(z\coloneqq bk_U-an_U\), and \(W\coloneqq U^{\oplus a}\oplus0^z\). Since \(k_U\ge Rn_U\), we have \(z\in\Z_{\ge0}\), and
		\begin{align*}
			\dim_{\F_{q^m}}(W)&=ak_U,\quad N=an_U+z=bk_U,\quad \rate(W)=a/b,\\
			n&\le an_U\le N=bk_U\le b(n+\ell)\le cn.
		\end{align*}

		For rank weight, \(\wR(x,y)\ge\max\{\wR(x),\wR(y)\}\) and \(\wR(x,0)=\wR(x)\). Indeed, \(\langle x_1,\ldots,x_n\rangle_{\F_q}\subseteq\langle x_1,\ldots,x_n,y_1,\ldots,y_\ell\rangle_{\F_q}\), and the same holds for \(y\). Thus, for nonzero component codes, \(\dR(C\oplus D)=\min\{\dR(C),\dR(D)\}\): the lower bound follows by projection, and \((x,0)\) or \((0,y)\), with \(x\) or \(y\) of minimum weight, attains it. Therefore
		\[
		\dR(W)=\dR(U)=
		\begin{cases}
			\dR(C),&C\ne0,\quad \dR(C)\le n\le \dR(D),\\
			\dR(D),&C=0,\quad U=0^n\oplus D,\ \dR(D)\ge n.
		\end{cases}
		\]

		Define \(u^\sharp\coloneqq(u,0^\ell,0^{(a-1)n_U},0^z)\in \F_{q^m}^N\). Only the first copy is translated: \(u^\sharp+W=((u+C)\oplus D)\oplus U^{\oplus(a-1)}\oplus0^z\). Projection onto the first \(n\) coordinates gives \(\wR(y)\ge\muR(u+C)\) for every \(y\in u^\sharp+W\). Choose \(x_*\in u+C\) with \(\wR(x_*)=\muR(u+C)\). Since \((x_*,0^{N-n})\in u^\sharp+W\),
		\[
		\muR(u+C)\le\muR(u^\sharp+W)\le\wR(x_*,0^{N-n})=\wR(x_*)=\muR(u+C).
		\]
		Finally, \(\dR(W)\ge\eta n\ge(\eta/c)N\). For \(\theta>0\), \(T\ge\theta n\ge(\theta/c)N\); for \(\theta\le0\), the threshold bound follows from \(T\ge0\). All operations use explicit linear algebra and direct sums.
	\end{proof}

	\begin{corollary}[Exact-rate Hamming-to-rank reduction]
		\label{body:exact-lift}
		Fix \(q\), a rational \(R\in(0,1)\), and \(\eps>0\). Given a nonzero \(C\le\F_q^{n_0}\) and a threshold \(t\), the preceding constructions produce, in deterministic polynomial time, a rank instance \((W,T)\) over \(\F_{q^m}\) with
		\[
		\begin{gathered}
			\rate(W)=R,\quad \dR(W)=L\cdot\dH(C),\quad T=Lt,\\
			N_1\le N\le c_RN_1,\quad m=O_{R,\eps}(N^{1+\eps}),
		\end{gathered}
		\]
		where \(L=n_0^{B-1}\) and \(N_1=n_0^B\) as in \Cref{body:repetition}. If \(\dH(C)\ge\eta_0n_0\) and \(t\ge\theta_0n_0\), then \(\dR(W)\ge(\eta_0/c_R)N\) and \(T\ge(\theta_0/c_R)N\).
	\end{corollary}

	\begin{proof}
		Apply \Cref{body:repetition} with \(\kappa=\kappa_R\). It gives distance \(L\cdot\dH(C)\) and \(m\ge\kappa_RN_1\), so \Cref{body:calibration} applies over the same field with input length \(N_1\). Keeping \(T\coloneqq Lt\), we obtain \(\dR(W)=L\cdot\dH(C)\), \(N_1\le N\le c_RN_1\), and \(m=O_{R,\eps}(N_1^{1+\eps})=O_{R,\eps}(N^{1+\eps})\). Since \(Ln_0=N_1\), for positive \(\eta_0\) and \(\theta_0\), respectively,
		\[
		\begin{aligned}
			\dR(W)&\ge L\eta_0n_0=\eta_0N_1\ge(\eta_0/c_R)N,\\
			T&\ge L\theta_0n_0=\theta_0N_1\ge(\theta_0/c_R)N.
		\end{aligned}
		\]
		A nonpositive constant gives the corresponding bound directly from \(\dR(W),T\ge0\).
	\end{proof}

	\subsection{Deterministic Gabidulin codes}
	\label{subsec:gabidulin}

	To change the rate of a nonzero code \(\cC\le \F_{q^m}^n\), we need an auxiliary code whose distance is at least \(\dR(\cC)\). Since rank weight is at most \(n\), \Cref{lem:elementary,lem:rank-directsum} give
	\[
	\dR(\cD)\ge n\ge\dR(\cC)\implies\dR(\cC\oplus\cD)=\min\{\dR(\cC),\dR(\cD)\}=\dR(\cC).
	\]
	Gabidulin codes provide these auxiliary codes explicitly and attain the largest rank distance for their dimension.

	An \(\F_q\)-linearized polynomial over \(\F_{q^m}\) has the form \(f(X)\coloneqq\sum_{j\ge0}a_jX^{q^j}\), with \(a_j\in \F_{q^m}\). Its evaluation map is \(\F_{q^m}\to \F_{q^m}\), \(x\mapsto f(x)\), and
	\[
	f(ux+vy)=uf(x)+vf(y),\quad x,y\in \F_{q^m},\ u,v\in\F_q.
	\]
	For \(f(X)\ne0\), its \(q\)-degree is the largest \(s\) with \(a_s\ne0\). The ordinary polynomial root bound gives
	\[
	\deg_X(f(X))=q^s\implies|\ker(x\mapsto f(x))|\le q^s\implies\dim_{\F_q}(\ker(x\mapsto f(x)))\le s.
	\]

	For an \(r\)-dimensional subspace \(W\le \F_{q^m}\), define \(A_W(X)\coloneqq\prod_{w\in W}(X-w)\). We verify that it is linearized of \(q\)-degree \(r\). The base case is \(A_{\{0\}}(X)=X\). Suppose the assertion holds for \(W\), and choose \(u\notin W\). Then \(A_W(u)\ne0\), and
	\begin{align*}
		A_{W\oplus\F_qu}(X)&=\prod_{c\in\F_q}\prod_{w\in W}(X-w-cu)\\
		&=\prod_{c\in\F_q}\bigl(A_W(X)-cA_W(u)\bigr)\\
		&=A_W(X)^q-A_W(u)^{q-1}A_W(X).
	\end{align*}
	The last polynomial is monic and linearized, of \(q\)-degree \(r+1\). Induction proves the assertion. Its roots are exactly \(W\), and any monic polynomial of degree \(q^r\) vanishing on all \(q^r\) elements of \(W\) equals this product. Thus \(A_W(X)\) is the unique monic linearized polynomial of \(q\)-degree \(r\) with kernel \(W\).

	If \(f(X)\) and \(g(X)\) have \(q\)-degrees \(s\) and \(r\), with leading coefficients \(a_s,b_r\ne0\), then \(g(f(X))\) is linearized and
	\[
	g(f(X))=b_ra_s^{q^r}X^{q^{r+s}}+\tx{lower-degree terms},\quad b_ra_s^{q^r}\ne0.
	\]
	In particular, its \(q\)-degree is \(r+s\).

	\begin{lemma}[Explicit Gabidulin code]
		\label{lem:gabidulin}
		Let \(1\le\ell\le m\). For every integer \(h\) with \(1\le h\le\ell\), one can construct deterministically, in time polynomial in \(m\), \(\ell\), and \(\log q\), an \(\F_{q^m}\)-linear code \(\cD\le \F_{q^m}^{\ell}\) with
		\[
		\dim_{\F_{q^m}}(\cD)=h, \quad \dR(\cD)=\ell-h+1.
		\]
	\end{lemma}

	\begin{proof}
		Choose \(\F_q\)-linearly independent \(g_1,\ldots,g_\ell\in \F_{q^m}\). In a polynomial basis \(\F_{q^m}=\F_q[\vartheta]\), take \(g_i\coloneqq\vartheta^{i-1}\); in any explicit representation, take the first \(\ell\) vectors of a known \(\F_q\)-basis. This is possible because \(\ell\le m\). Define \(\cD\) as the image of the \(\F_{q^m}\)-linear evaluation map
		\[
		\F_{q^m}^h\longrightarrow \F_{q^m}^\ell,\quad(a_0,\ldots,a_{h-1})\longmapsto(f(g_1),\ldots,f(g_\ell)),\quad f(X)\coloneqq\sum_{j=0}^{h-1}a_jX^{q^j}.
		\]
		A generator matrix is
		\[
		\bigl(g_i^{q^j}\bigr)_{0\le j<h,\,1\le i\le\ell}\in \F_{q^m}^{h\times\ell}.
		\]
		To show injectivity, suppose a coefficient vector evaluates to zero. If \(f(X)\ne0\), then
		\[
		\langle g_1,\ldots,g_\ell\rangle_{\F_q}\subseteq\ker(x\mapsto f(x))\implies\ell\le\dim_{\F_q}(\ker(x\mapsto f(x)))\le h-1<\ell.
		\]
		This is a contradiction. Hence \(f(X)=0\), all coefficients vanish, and \(\dim_{\F_{q^m}}(\cD)=h\).

		For the distance, let \(f(X)\ne0\) have \(q\)-degree \(s\le h-1\), and define
		\[
		y\coloneqq(f(g_1),\ldots,f(g_\ell)),\quad W\coloneqq\langle f(g_1),\ldots,f(g_\ell)\rangle_{\F_q},\quad r\coloneqq\dim_{\F_q}(W)=\wR(y).
		\]
		Suppose \(r\le\ell-h\). The map \(\F_{q^m}\to \F_{q^m}\), \(x\mapsto f(x)\), satisfies
		\begin{align*}
			\dim_{\F_q}(\{f(x):x\in \F_{q^m}\})&=m-\dim_{\F_q}(\ker(x\mapsto f(x)))\\
			&\ge m-s\ge m-(h-1)\ge\ell-h+1>r.
		\end{align*}
		Thus there is \(w\in \F_{q^m}\) with \(z\coloneqq f(w)\notin W\). By the defining property of \(A_W(X)\),
		\[
		A_W(f(w))=A_W(z)\ne0,\quad\tx{so}\quad A_W(f(X))\ne0.
		\]
		This composition is linearized of \(q\)-degree at most
		\[
		r+s\le(\ell-h)+(h-1)=\ell-1.
		\]
		But \(A_W(f(g_i))=0\) for every \(i\in[\ell]\), and hence
		\[
		\langle g_1,\ldots,g_\ell\rangle_{\F_q}\subseteq\ker(x\mapsto A_W(f(x)))\implies\ell\le\dim_{\F_q}(\ker(x\mapsto A_W(f(x))))\le\ell-1.
		\]
		This contradiction proves \(\wR(y)\ge\ell-h+1\) for every nonzero codeword.

		For the matching upper bound, rank weight is at most Hamming weight over \(\F_{q^m}\), by \Cref{lem:elementary}, and the Hamming Singleton bound gives
		\[
		\ell-h+1\le\dR(\cD)\le\dH(\cD)\le\ell-h+1.
		\]
		The generator has \(h\ell\) entries in \(\F_{q^m}\), each represented by \(O(m\cdot\log q)\) bits. Computing its successive Frobenius powers and the evaluation basis takes time polynomial in \(m,\ell,\log q\).
	\end{proof}

	\section{The Hamming source}
	\label{sec:body-source}

	We need both ordinary and distinguished-coordinate Hamming instances in which every nonzero word has weight proportional to the length, and the YES and NO cases have separated weight bounds. We use the balanced-code construction and the distinguished coordinate of~\cite{BGR25}. The quadratic reduction and the balanced map are stated in~\Cref{prop:quadeq,lem:balanced}. The following weight bound gives the required separation.

	\begin{lemma}[Weight separation]
		\label{body:weight}
		Let \(D>0\) and \(\eps\ge0\), and let \(G\colon\Fq^p\to\Fq^M\) be an injective linear map, represented by \(G\in\Fq^{M\times p}\), with \(D\le\|Gx\|_0\le(1+\eps)D\) for every \(x\ne0\). Then \(\|GXG^{\transpose}\|_0\ge D^2\) for every nonzero \(X\in\Fq^{p\times p}\). If \(\rank X=1\), this weight is at most \((1+\eps)^2D^2\); if \(\rank X\ge2\), it is at least \((1+1/q)D^2\).
	\end{lemma}

	\begin{proof}
		Put \(U\coloneqq G(\Fq^p)\) and \(d\coloneqq\dH(U)\ge D\). For linearly independent \(u,v\in U\), each coordinate in \(\supp(u)\cup\supp(v)\) is nonzero in exactly \(q^2-q\) of the \(q^2-1\) words \(au+bv\), \((a,b)\in\Fq^2\setminus\{(0,0)\}\). Every such word has weight at least \(d\), so
		\[
		(q^2-q)|\supp(u)\cup\supp(v)|=\sum_{(a,b)\ne(0,0)}\|au+bv\|_0\ge(q^2-1)d.
		\]
		Consequently, \(|\supp(u)\cup\supp(v)|\ge(q^2-1)d/(q^2-q)=(1+1/q)d\).

		Put \(Y\coloneqq GXG^{\transpose}\). Its columns and transposed rows lie in \(U\). Choose a left inverse \(P\in\Fq^{p\times M}\) of \(G\). Then \(PG=I_p,\quad X=PYP^{\transpose},\quad \rank X\le\rank Y\le\rank X.\) If \(Y_{*,j}\ne0\), every row indexed by \(\supp(Y_{*,j})\) is nonzero, and hence
		\[
		\|Y\|_0=\sum_i\|Y_{i,*}\|_0\ge d\,|\supp(Y_{*,j})|\ge d^2\ge D^2.
		\]
		If \(\rank Y\ge2\), choose two independent columns \(Y_{*,j},Y_{*,k}\). The same row sum gives
		\[
		\|Y\|_0\ge d\,|\supp(Y_{*,j})\cup\supp(Y_{*,k})|\ge(1+1/q)d^2\ge(1+1/q)D^2.
		\]
		Finally, if \(\rank X=1\), write \(X=uv^{\transpose}\) with nonzero \(u,v\in\Fq^p\). Then \(Y=(Gu)(Gv)^{\transpose}\), and \(\|Y\|_0=\|Gu\|_0\|Gv\|_0\le(1+\eps)^2D^2\).
	\end{proof}

	\begin{lemma}[Ordinary and distinguished-coordinate sources]
		\label{body:source}
		Fix a prime power \(q\) and a rational \(\Lambda>1\). There are deterministic polynomial-time reductions from \(3\)SAT of the following two forms, with positive constants depending only on \(q,\Lambda\).
		\begin{enumerate}
			\item The ordinary reduction outputs a nonzero code \(C\le\Fq^{n_o}\) and an integer \(T_o\) with \(\dH(C)\ge\eta_o n_o\) and \(\theta_o n_o\le T_o\le n_o\). In the YES case \(\dH(C)\le T_o\); in the NO case \(\dH(C)>\Lambda T_o\).
			\item The distinguished-coordinate reduction outputs \(V\le\Fq^{n_m}\), a coordinate \(j\), and an integer \(T_m\) with \(\theta_m n_m\le T_m\le n_m\), and \(\|v\|_0\ge\eta_m n_m\) for every \(v\in V\setminus\{0\}\). In the YES case there is \(v\in V\) with \(v_j=1\) and \(\|v\|_0\le T_m\); in the NO case \(\|v\|_0>\Lambda T_m\) for every \(v\in V\setminus\{0\}\). The code \(V\) may be zero in the NO case.
		\end{enumerate}
		The lengths and constants of the two reductions need not coincide.
	\end{lemma}

	\begin{proof}
		By \Cref{prop:quadeq}, reduce \(3\)SAT to equations \(Q_i(ww^{\transpose})=0\) over \(\Fq\), where the \(Q_i\) are linear forms on \(p\times p\) matrices. In the YES case there is a solution with \(w_z=1\); in the NO case there is no nonzero solution.

		Set \(\xi_q\coloneqq1/(18q)\), \(\eps_q\coloneqq1/(9q)\), \(\delta_q\coloneqq(1-\xi_q)(1-1/q)\), \(A_q\coloneqq1+1/(3q)\), and \(B_q\coloneqq1+1/q\). By \Cref{lem:balanced}, construct \(G\colon\Fq^p\to\Fq^M\) with \(D\coloneqq\delta_qM\) and \(D\le\|Gx\|_0\le(1+\eps_q)D\) for \(x\ne0\). Here \(M\) is bounded by a polynomial in \(p\) and tends to infinity with \(p\). Also, \((1+\eps_q)^2\le A_q<B_q\).

		Solve the linear system defining \(S\coloneqq\{X\in\Fq^{p\times p}:X=X^{\transpose},\quad Q_i(X)=0\ \tx{for all }i\}.\) Reading a matrix row by row as a vector of length \(M^2\), form the image codes \(V_0\coloneqq\phi_0(S)\) and \(V_1\coloneqq\phi_1(S)\) of the maps
		\[
		\begin{aligned}
			\phi_0:S&\longrightarrow\Fq^{M^2},&X&\longmapsto GXG^{\transpose},\\
			\phi_1:S&\longrightarrow\Fq^{M^2+1},&X&\longmapsto(GXG^{\transpose},X_{zz}).
		\end{aligned}
		\]
		Both maps are injective, since \(X=P(GXG^{\transpose})P^{\transpose}\) for a left inverse \(P\) of \(G\). By \Cref{body:weight}, \(\|\phi_b(X)\|_0\ge D^2\) for \(b\in\{0,1\}\) and \(X\in S\setminus\{0\}\).

		In a YES instance, choose a solution \(w\) with \(w_z=1\). Then \(X\coloneqq ww^{\transpose}\in S\), \(X_{zz}=1\), \(\|\phi_0(X)\|_0\le A_qD^2\), and \(\|\phi_1(X)\|_0\le A_qD^2+1\).

		In a NO instance, a nonzero rank-one \(X\in S\) would satisfy \(X=uv^{\transpose}=vu^{\transpose}\) for nonzero \(u,v\in\Fq^p\). Symmetry gives \(v=\lambda u\) for some \(\lambda\ne0\), since \(\langle u\rangle_{\F_q}=\operatorname{col}X=\operatorname{col}X^{\transpose}=\langle v\rangle_{\F_q}\), and hence \(0=Q_i(X)=\lambda Q_i(uu^{\transpose})\quad\implies\quad Q_i(uu^{\transpose})=0\quad\tx{for all }i,\) contradicting the quadratic NO case. Thus \(X\in S\setminus\{0\}\) implies \(\rank X\ge2\), and \(\|\phi_b(X)\|_0\ge B_qD^2\) for \(b\in\{0,1\}\).

		Choose a fixed integer \(a\ge1\) with \((B_q/A_q)^a>\Lambda\). For \(b\in\{0,1\}\), put
		\[
		W_b\coloneqq V_b^{\otimes a},\quad\ell_b\coloneqq(M^2+b)^a,\quad t_b\coloneqq\min\{\ell_b,\lceil(A_qD^2+b)^a\rceil\}.
		\]
		If \(V_b=0\), then \(W_b=0\), and there are no nonzero words to consider. Otherwise, minimum distance multiplies under tensor products by~\Cref{lem:tensor-full}. Thus, for every \(0\ne v\in W_b\),
		\[
		\|v\|_0\ge D^{2a},\quad\tx{and in a NO instance,}\quad\|v\|_0\ge(B_qD^2)^a.
		\]
		In a YES instance, the previously constructed witness \(v_b\in V_b\) satisfies
		\[
		\|v_b^{\otimes a}\|_0=\|v_b\|_0^a\le\min\{\ell_b,(A_qD^2+b)^a\}\le t_b.
		\]
		For \(b=1\), its distinguished tensor coordinate satisfies \((v_1^{\otimes a})_{(M^2+1,\ldots,M^2+1)}=1^a=1\).

		Since \(D\to\infty\) with \(p\), we have
		\[
		\frac{(B_qD^2)^a}{(A_qD^2+1)^a+1}\longrightarrow(B_q/A_q)^a>\Lambda.
		\]
		Hence, for all sufficiently large \(p\),
		\[
		(B_qD^2)^a>\Lambda\bigl((A_qD^2+1)^a+1\bigr)\ge\Lambda t_b.
		\]
		The length and threshold satisfy
		\[
		\delta_q^{2a}M^{2a}=D^{2a}\le t_b\le\ell_b\le2^aM^{2a}.
		\]
		The distinguished-coordinate output is \(V\coloneqq W_1\), \(n_m\coloneqq\ell_1\), \(T_m\coloneqq t_1\), with distinguished coordinate \((M^2+1,\ldots,M^2+1)\) and \(\eta_m=\theta_m\coloneqq\delta_q^{2a}/2^a\).

		For the ordinary output, put
		\[
		h\coloneqq\lfloor\Lambda t_0\rfloor+1,\quad C\coloneqq W_0\oplus\Rep_h,\quad n_o\coloneqq\ell_0+h,\quad T_o\coloneqq t_0,
		\]
		where \(\Rep_h\coloneqq\{\lambda1^h:\lambda\in\Fq\}\). This code is nonzero even when \(W_0=0\). With the convention \(\dH(0)=+\infty\),
		\[
		\dH(C)=\min\{\dH(W_0),h\},\quad h>\Lambda t_0\ge D^{2a}.
		\]
		The YES witness lies in \(W_0\oplus0^h\), so \(\dH(C)\le t_0\) in the YES case. In the NO case both entries of the minimum exceed \(\Lambda t_0\). In either case \(\dH(C)\ge D^{2a}\). Moreover,
		\begin{align*}
			n_o&=M^{2a}+h\le M^{2a}+\Lambda t_0+1\le(\Lambda+2)M^{2a},\\
			\dH(C),\ T_o&\ge D^{2a}=\delta_q^{2a}M^{2a}\ge\frac{\delta_q^{2a}}{\Lambda+2}n_o.
		\end{align*}
		Thus \(\eta_o=\theta_o\coloneqq\delta_q^{2a}/(\Lambda+2)\) suffice.

		For the bounded values of \(p\) excluded above, decide the quadratic system by exhaustive search. The ordinary reduction outputs
		\[
		(C,T_o)\coloneqq
		\begin{cases}
			(\Rep_1,1),&\tx{on YES},\\
			(\Rep_{\lfloor\Lambda\rfloor+1},1),&\tx{on NO}.
		\end{cases}
		\]
		These outputs have relative distance \(1\), and the ratio of threshold to length is at least \((\lfloor\Lambda\rfloor+1)^{-1}\). Their distances are \(1\le T_o\) and \(\lfloor\Lambda\rfloor+1>\Lambda T_o\), respectively. The distinguished-coordinate reduction outputs
		\[
		(V,j,T_m)\coloneqq
		\begin{cases}
			(\Fq,1,1),&\tx{on YES},\\
			(\{0\},1,1),&\tx{on NO}.
		\end{cases}
		\]
		The YES word is \(1\), the NO code has no nonzero word, and both thresholds equal their length. Decrease the constants \(\eta_o,\theta_o,\eta_m,\theta_m\), if necessary, to include these outputs. All thresholds are computed from \(D\), not from minimum distances. Linear algebra, the balanced map, and the fixed tensor power take polynomial time.

	\end{proof}

	\subsection{Imported ingredients}

	The following proposition is \cite[Proposition~5]{BGR25}, applied after translating a \(3\)SAT formula into a Boolean circuit.

	\begin{proposition}[Homogeneous quadratic equations]
		\label{prop:quadeq}
		For every fixed finite field \(\F_q\), there is a deterministic polynomial-time reduction from \(3\)SAT to linear forms \(Q_1,\ldots,Q_\ell\) on \(\F_q^{p\times p}\) and a distinguished coordinate \(z\in[p]\) such that:
		\begin{itemize}
			\item (YES) there is \(w\in\F_q^p\) with \(w_z=1\) and \(Q_i(ww^{\transpose})=0\) for every \(i\in[\ell]\);
			\item (NO) there is no nonzero \(w\in\F_q^p\) with \(Q_i(ww^{\transpose})=0\) for every \(i\in[\ell]\).
		\end{itemize}
	\end{proposition}

	We also use the explicit construction in \cite[Lemma~11]{BGR25}. The bound \(D\) is known from the output length, so we can compute the thresholds above without computing a minimum distance.

	\begin{lemma}[Explicit balanced map]
		\label{lem:balanced}
		For every fixed prime power \(q\), every fixed rational \(\xi\in(0,1/2)\), and every input dimension \(n\ge1\), there is a deterministic polynomial-time construction of a linear map \(G:\F_q^n\to\F_q^N\) and a computable rational number
		\[
		D\coloneqq(1-\xi)\Bigl(1-\tfrac1q\Bigr)N
		\]
		such that \(N\le(qn/\xi)^2\) and every nonzero \(x\in\F_q^n\) satisfies
		\[
		D\le\|Gx\|_0\le\Bigl(1-\tfrac1q\Bigr)N\le(1+2\xi)D.
		\]
		In particular, \(G\) is injective.
	\end{lemma}

	\section{Deterministic rank-metric hardness\\
		{\small Proof of Theorem~\ref{thm:intro-rmdp}, Theorem~\ref{thm:intro-rncp}}}
	\label{body:hardness-section}

	We now combine the Hamming reductions, the distance-preserving lift, and the rate calibration to prove the two hardness results.

	\begin{proof}[Proof of \Cref{thm:intro-rmdp,thm:intro-rncp}]
		Fix a rational \(R\in(0,1)\), choose a rational \(\Lambda>\Gamma\), and let \(\eta_0,\theta_0\) be the smaller constants in the two outputs of \Cref{body:source}. For minimum distance, apply \Cref{body:exact-lift} to the ordinary output \((C,t)\). It gives \(\dR(W)=L\cdot\dH(C)\) and \(T=Lt\), so \(\dH(C)\le t\implies\dR(W)\le T\) and \(\dH(C)>\Lambda t\implies\dR(W)>\Lambda T>\Gamma T\). Also \(\rate(W)=R\), \(\dR(W)\ge(\eta_0/c_R)N\), \(T\ge(\theta_0/c_R)N\), and \(m=O_{R,\eps}(N^{1+\eps})\).

		For decoding, take the distinguished-coordinate source \((V,j,t)\) of length \(n_0\). For a basis \(v_1,\ldots,v_k\) of \(V\), Gaussian elimination tests whether \(\sum_{i=1}^k\lambda_i(v_i)_j=1\) has a solution \((\lambda_1,\ldots,\lambda_k)\in\F_q^k\), equivalently whether the slice \(\{v\in V:v_j=1\}\) is nonempty. Suppose first that it is nonempty, so \(V\ne0\).

		Apply \Cref{body:repetition} with \(\kappa=\kappa_R\) to obtain \(\cV=D_\alpha(V^{(L)})_{\F_{q^m}}\le \F_{q^m}^{N_1}\). Let \(j_\star\) be one of the repeated copies of \(j\), and put \(\cA\coloneqq\{x\in\cV:x_{j_\star}=\alpha_{j_\star}\}\), \(\cC_0\coloneqq\{x\in\cV:x_{j_\star}=0\}\), and \(T\coloneqq Lt\). Choose any \(v_0\in V\) with \((v_0)_j=1\), and put \(x_0\coloneqq D_\alpha(v_0^{(L)})\). Then \((x_0)_{j_\star}=\alpha_{j_\star}\) and \(\cA=x_0+\cC_0\).

		In a YES instance, choose the witness \(v\in V\) with \(v_j=1\) and \(\|v\|_0\le t\). Since every \(\alpha_i\ne0\), its image belongs to \(\cA\) and satisfies
		\[
		\wR(D_\alpha(v^{(L)}))\le\|D_\alpha(v^{(L)})\|_0=L\|v\|_0\le Lt=T.
		\]
		In a NO instance, every \(x\in\cA\) has \(x_{j_\star}=\alpha_{j_\star}\ne0\), and hence
		\[
		\wR(x)\ge\dR(\cV)=L\cdot\dH(V)>\Lambda Lt=\Lambda T.
		\]
		Further, for every \(x\in\cC_0\setminus\{0\}\), \(\wR(x)\ge\dR(\cV)\ge L\eta_0n_0=\eta_0N_1\). Since \(m\ge\kappa_RN_1\), apply \Cref{body:calibration} to \(\cA\), retaining threshold \(T\). If \(\cC_0=0\), its zero-code case gives \(\dR(W)\ge N_1\ge\eta_0N_1\); otherwise its distance formula gives the same lower bound. Thus
		\[
		\begin{aligned}
			\muR(x_0^\sharp+W)&=\muR(\cA),\quad \rate(W)=R,\quad N_1\le N\le c_RN_1,\\
			\dR(W)&\ge\eta_0N_1\ge(\eta_0/c_R)N,\quad T\ge\theta_0N_1\ge(\theta_0/c_R)N.
		\end{aligned}
		\]
		Also \(m=O_{R,\eps}(N_1^{1+\eps})=O_{R,\eps}(N^{1+\eps})\).

		If \(\{v\in V:v_j=1\}\) is empty, the source instance is NO. Fix an integer \(D>\Gamma\) and a degree \(m_*\ge\max\{2,D,\lceil\kappa_RD\rceil\}\). Choose \(a_1,\ldots,a_D\in \F_{q^{m_*}}\) linearly independent over \(\F_q\). For \(a\coloneqq(a_1,\ldots,a_D)\), \(\wR(a)=\dim_{\F_q}(\langle a_1,\ldots,a_D\rangle)=D>\Gamma.\) Calibrate \(\{a\}=a+0\) with threshold \(1\), obtaining a fixed output \(u_*+W_*\subseteq\F_{q^{m_*}}^{N_*}\) with \(\muR(u_*+W_*)=D\), \(\rate(W_*)=R\), and \(\dR(W_*)\ge D\).

		Take \(\eta\coloneqq\min\{\eta_0/c_R,D/N_*\}>0\) and \(\theta\coloneqq\min\{\theta_0/c_R,1/N_*\}>0\). If the nonempty branch has degree bound \(m\le c_{\tx{\normalfont deg}}N^{1+\eps}\), replace \(c_{\tx{\normalfont deg}}\) by \(\max\{c_{\tx{\normalfont deg}},m_*\}\) to cover the fixed output.

		Finally, for any output \(u+W\subseteq \F_{q^m}^N\), the nearest-codeword instance \((W,u,T)\) has optimum \(\dist(u,W)=\muR(u+W)\). Compute a full-row-rank parity-check matrix \(H\in \F_{q^m}^{(N-\dim_{\F_{q^m}}(W))\times N}\) with \(\ker H=W\), and put \(s\coloneqq Hu^{\transpose}\). Then
		\[
		\{e\in \F_{q^m}^N:He^{\transpose}=s\}=u+W,\quad\{e\in \F_{q^m}^N:He^{\transpose}=0\}=W.
		\]
		Thus the syndrome instance is consistent and has the same optimum and kernel as the affine instance.

		Field construction, elimination, repetition, lifting, and calibration all take deterministic polynomial time for the fixed parameters, by~\Cref{lem:fields,body:repetition,body:calibration}. The constants \(\eta,\theta>0\) depend only on \(q,R,\Gamma\).
	\end{proof}

	\clearpage

	\printbibliography
\end{document}